\documentclass[conf]{new-aiaa}
\usepackage[utf8]{inputenc}

\usepackage{graphicx}
\usepackage{amsmath}
\usepackage[version=4]{mhchem}
\usepackage{siunitx}
\usepackage{longtable,tabularx}
\usepackage{ulem}

\usepackage{xcolor}
\usepackage[caption=false,font=footnotesize]{subfig}
\usepackage{floatrow}
\usepackage[font=normalsize]{caption}
\usepackage{epsfig} % for postscript graphics files
\usepackage{algorithm}
\usepackage{algpseudocode}

\usepackage{bm}
\usepackage{dsfont}

\usepackage{ntheorem}\theoremseparator{.}
\theoremseparator{.}
\theoremindent\parindent
\theoremseparator{.}
\theoremseparator{.}
\theoremseparator{.}
\theoremseparator{.}
\theoremseparator{.}

\usepackage{comment}
\usepackage{amsfonts} 
\usepackage{scrextend}
\usepackage{ragged2e}

\usepackage{epsfig}
\newcommand{\R}{\mathbb{R}}

\newcommand{\T}{\sf T}

\newcommand{\N}{\mathbb{N}}

\title{
Constrained Control for Autonomous Spacecraft Rendezvous: Learning-Based Time Shift Governor
}

\author{Taehyeun Kim\footnote{Ph.D. candidate, Aerospace Engineering, 1320 Beal Ave, Ann Arbor, MI 48109}${}^{,1}$,  Robin Inho Kee\footnote{Master's student, Department of Mechanical Engineering, University of Michigan}${}^{,1}$, Ilya Kolmanovsky\footnote{Professor, Aerospace Engineering, 3038 FXB, 1320 Beal Ave, Ann Arbor, MI 48109, AIAA Associate Fellow.}, and Anouck Girard\footnote{Professor, Robotics and Aerospace Engineering, 3264 FMCRB,
2505 Hayward Street, Ann Arbor, MI 48109, AIAA Associate Fellow.\\ 
  \indent \hspace{\parindent} \( {}^{1} \)These authors contributed equally to this work} 
  }
\affil{Department of Aerospace Engineering, University of Michigan, Ann Arbor, 48109 MI, USA}

\begin{document}

\maketitle

\begin{abstract}
This paper develops a Time Shift Governor (TSG)-based control scheme to enforce constraints during rendezvous and docking (RD) missions in the setting of the Two-Body problem. As an add-on scheme to the nominal closed-loop system, the TSG generates a time-shifted Chief spacecraft trajectory as a target reference for the Deputy spacecraft. This modification of the commanded reference trajectory ensures that constraints are enforced while the time shift is reduced to zero to effect the rendezvous. Our approach to TSG implementation integrates an LSTM neural network which approximates the time shift parameter as a function of a sequence of past Deputy and Chief spacecraft states. This LSTM neural network is trained offline from simulation data. We report simulation results for RD missions in the Low Earth Orbit (LEO) and on the Molniya orbit to demonstrate the effectiveness of the proposed control scheme. The proposed scheme reduces the time to compute the time shift parameter in most of the scenarios and successfully completes rendezvous missions. 
\end{abstract}

%\section{Nomenclature} \label{sec:nomenclature}
%\input{./sections/nomenclature.tex}

%\section{Introduction} \label{sec:intro}
%\input{./sections/intro.tex}

\section{Introduction} \label{sec:intro}
Spacecraft proximity operations (SPO) involve two spacecraft maneuvering near each other in space, e.g., to perform rendezvous and docking (RD). The primary spacecraft (referred to as the Chief) maintains a nominal orbit passively or actively, while the other spacecraft (referred to as the Deputy) is actively controlled to perform the RD mission~\cite{hartley2015tutorial} while satisfying constraints on thrust magnitude, on the line of sight (LoS) constraint of the docking port and on relative velocity.

% Is the RD mission important? - yes, many missions are performed and planned to be done in the future.
%The RD mission plays a crucial role in many space programs. The RD missions can be conducted manually, as was done for the Skylab mission in 1973, or autonomously, as was done for the Orbital Express mission in 2007. For example, since 2014, Northrop Grumman has conducted a series of commercial resupply services to the ISS. Also, since 2020, SpaceX has launched spacecraft docking missions and has provided supply service to NASA and a commercial company in SpaceX Crew missions, and Axiom Mission 1. The private sector has shown that they can provide the RD service to NASA and private companies, considering multiple test and operational crewed missions to transport astronauts and supplies in the commercial space program. 

Many spacecraft RD missions were conducted near and beyond the Earth. The first spacecraft docking mission was a part of the Gemini 8 mission, in which astronauts manually performed the docking with the target vehicle. The following RD missions were also conducted, as was done for the Apollo mission in 1969 and the Skylab mission in 1973. As docking techniques were cultivated through former missions, docking missions became complicated and required an autonomous system to ensure safety and reduce costs. The first autonomous spacecraft docking mission was performed as a part of the Orbital Express mission in 2007. As examples in private sectors, Northrop Grumman then began to provide commercial resupply services to the ISS in 2014, and SpaceX launched spacecraft docking missions for supply service to NASA and for Axiom Space in 2020. Thus, the RD techniques played a crucial role in space missions while satisfying the growing interest of the space community in complicated missions.

Various control approaches for RD have been studied. In particular, the artificial potential function (APF) methods have been applied to ensure safety to address in RD operations~\cite{dong2017safety, dong2018dual}. However, the use of APFs to handle multiple constraints and simultaneous state and control constraints can be not straightforward. Model Predictive Control (MPC) based on the Clohessy-Wiltshire-Hill (CWH) linearized relative motion model has been considered in~\cite{2015weiss}. However, the utilization of MPC requires addressing the computational challenges inherent in solving a discrete-time optimal control problem online. The integration of genetic algorithms (GA) with fuzzy logic controllers (FLC) has been studied in~\cite{karr1997genetic}. However, this approach requires a subjective selection of fuzzy membership functions and lacks formal constraint enforcement guarantees. 
% Robust hybrid supervisory control to deal with different constraints and tasks to perform on docking phases.~\cite{malladi2016robust}

% What are current learning approaches in RVD? - RL but has limitations
The use of machine learning for spacecraft control has been of growing interest~\cite{huang2021spacecraft}. Deep reinforcement learning (DRL), in particular, has been considered for autonomous guidance and proximity operation~\cite{wang2020autonomous, hovell2021deep, federici2021deep, qu2022spacecraft, broida2019spacecraft}. 
Guaranteeing stability and convergence with DRL can, however, be challenging~\cite{xu2014reinforcement, tipaldi2022reinforcement}.

% We are introducing TSG approach with learning to overcome the formal learning approaches in RVD
% WHY TSGs? Add-on method, handles constraints, elegant implementations for s/c formations. WHY THIS MAY NOT WORK HERE? Computation issues for more complicated scenarios (3-4 body problems, elliptic orbits). 
% Existing learning methods face challenges, particularly in dealing with nonlinear dynamics and constraint satisfaction for space missions.
The Time Shift Governor (TSG) has been considered in ~\cite{frey2016time, frey2017parameter, kim2023time, kim2024time} to enforce constraints by time shifting the target trajectory of the Chief spacecraft and adjusting the time shift to achieve safe RD. The TSG, which is a variant of the parameter governor, is an add-on scheme used for managing constraints without redesigning the closed-loop system. It has been developed for scenarios involving circular Earth orbits in the Two-Body problem setting and more complex scenarios like halo orbits in the Circular Restricted Three-Body Problem (CR3BP) setting~\cite{frey2016time, kim2023time}. Unlike general nonlinear model predictive controllers that solve high-dimensional optimization problems, TSG reduces the optimization problem to optimizing a single time shift parameter.

In this paper, we consider an imitation learning motivated approach to the implementation of TSG which involves the use of a Long Short Term Memory (LSTM) network~\cite{han2019review, lara2021experimental} to approximate the optimal time shift.  The neural network is trained using data from offline simulation of the closed-loop system with TSG and then deployed for the onboard use to compute the time shift parameter without requiring iterative optimization. Notably, we find that by using a sequence of past states as the input to LSTM we are able to improve approximation accuracy with LSTM as compared to only using the current state.  We also propose a hybrid variant of the algorithm where the TSG online optimization involving bisections is only performed if the time shift parameter generated by the LSTM neural network is not safe due to approximation errors. 

% Paper organization
This paper is organized as follows: In Section II, we summarize the coordinate systems, spacecraft translational dynamics model, the nominal controller, as well as the constraints considered during the RD mission. Then, in Section III, we introduce the TSG to enforce the constraints. Section IV introduces our Learning-based TSG (L-TSG), which integrates an LSTM network with a phase-adaptive sliding window and a custom loss function, further enhanced by a hybrid prediction algorithm. Section V reports the numerical results, demonstrating the efficacy of our approach. Finally, Section VI concludes the paper with a summary of contributions and potential directions for further research.

\section{Problem Formulation} \label{sec:problem_formulation}
% In this work, we consider a spacecraft docking mission in an elliptic orbit around the Earth subject to multiple mission-specific constraints. During the docking mission, the secondary (Deputy) spacecraft locates near the primary (Chief) spacecraft. We use subscript \(c\) and \(d\) to denote the Chief spacecraft and the Deputy spacecraft, respectively, and we use the subscript \(i\) to denote spacecraft that can either be the Chief or the Deputy. 

In our study, we consider a docking mission involving the Deputy and the Chief spacecraft on orbit, with mission-specific constraints. 
% Throughout this mission, the secondary spacecraft, referred to as the Deputy, is positioned in proximity to the primary spacecraft, known as the Chief. 
% Specifically, given a rendezvous and docking (RD) problem on elliptic orbits with spacecraft dynamics, the aim is to find an optimal and computationally efficient parameter governor that satisfies mission constraints while accomplishing the mission without failure.
% Specifically, given a rendezvous and docking (RD) problem with two spacecraft on Low Earth Orbits (LEO) and the Molniya orbit, the goal is to find a trajectory of the optimal time shift parameters such that the Deputy spacecraft achieves the Chief spacecraft satisfying the mission constraints. 
This section will detail the coordinate systems, dynamic models, control strategies, and constraints that define our approach to the RD mission. We assign the subscripts \(c\) and \(d\) to denote the Chief and Deputy spacecraft respectively, while the subscript \(i\) is used to denote a spacecraft that could be either the Chief or the Deputy.

\subsection{Coordinate systems}\label{subsec2}
\begin{figure}[htbp]
\centering
\includegraphics[width=0.8\textwidth]{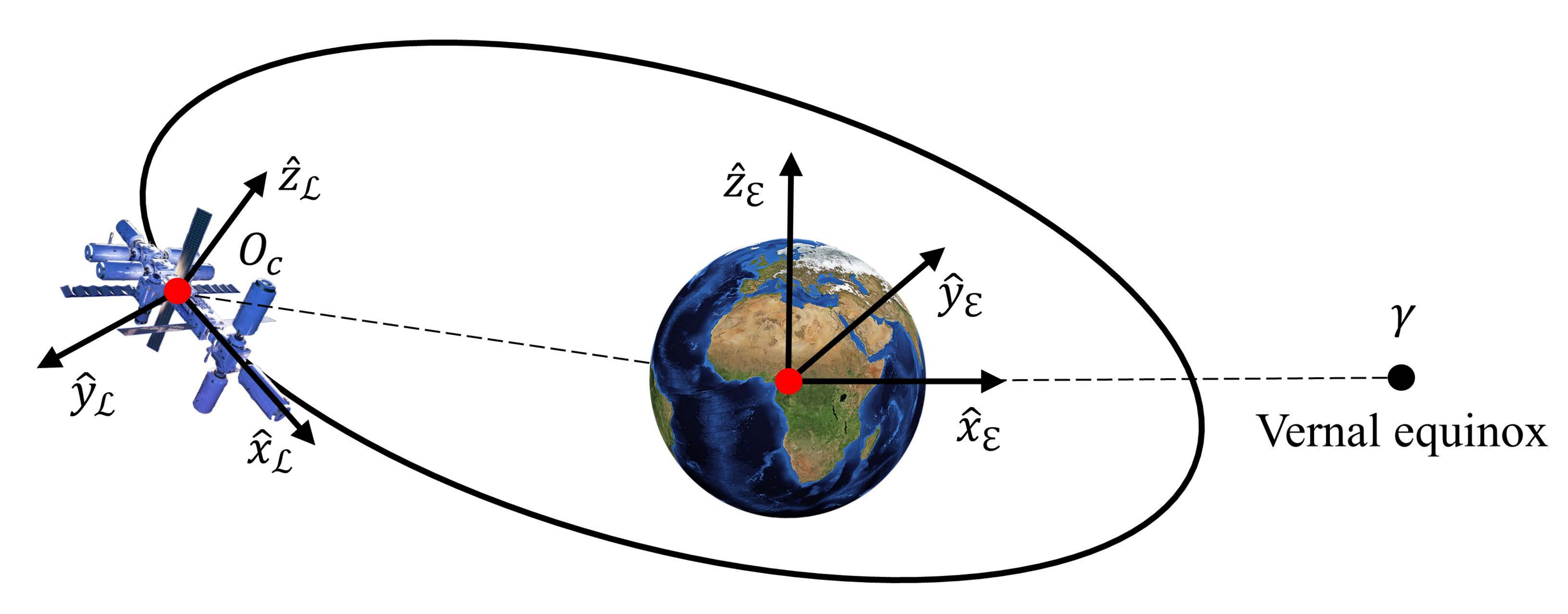}
\caption{ECI and VNB coordinate systems.}
\label{fig:eci_vnb}
\end{figure}

In this work, two right-handed coordinate systems are employed to describe the spacecraft equations of motion and relative motion, as shown in Fig .~\ref{fig:eci_vnb}. The origin of the Earth-centered inertial (ECI) frame, \( \mathcal{E} : \{\hat{x}_{\mathcal{E}}, \hat{y}_{\mathcal{E}}, \hat{z}_{\mathcal{E}} \} \), is at the center of the Earth $O_{\mathcal{E}}$, with the $\hat{x}$-axis pointing towards the vernal equinox—the direction from the Earth to the Sun at noon on the day of the spring equinox, the $\hat{z}$-axis pointing towards the Earth's rotational axis, and the $\hat{y}$-axis completing the right-handed system.

The local Velocity-Normal-Binormal (VNB) frame,  \( \mathcal{L} : \{O_{c}, \hat{x}_{\mathcal{L}}, \hat{y}_{\mathcal{L}}, \hat{z}_{\mathcal{L}}\} \), is also defined to describe the relative motion of objects with respect to the Chief spacecraft. Unlike the ECI frame, the VNB frame is a rotating and accelerating coordinate system and has its origin at the center of mass of the Chief spacecraft $O_{c}$ with three orthonormal vectors which are defined as
\begin{equation}
    \hat{x}_{\mathcal{L}} = \frac{v(X_{c})}{\| v(X_{c}) \|},\; \hat{y}_{\mathcal{L}} = \frac{p(X_{c}) \times \hat{x}_{\mathcal{L}}}{\| p(X_{c}) \times \hat{x}_{\mathcal{L}} \|} ,\; \hat{z}_{\mathcal{L}}=\hat{x}_{\mathcal{L}} \times \hat{y}_{\mathcal{L}},
\end{equation}
where $p(\cdot),v(\cdot): \R^{6}\to \R^{3}$ are functions that map from a state to position and velocity, respectively, expressed in the ECI frame.

\subsection{Dynamics}\label{subsec2}
The equations of motion of spacecraft are given by
\begin{equation}
\label{eq:equation_of_motion}
\dot{X}_i = f(t, X_i(t), u_i(t)),
\end{equation}
where \(X_i = [x_i, y_i, z_i, \dot{x}_i, \dot{y}_i, \dot{z}_i]^{\sf T},\) for $i\in \{c,d\}$, and \(u_{i} = [u_{1,i}, u_{2,i}, u_{3,i}]^{\sf T}\) denote the spacecraft state and the control input to the spacecraft, expressed in the ECI frame. Note that the Chief spacecraft maintains its nominal orbit without using control input, i.e., $u_{c}=0$, while the Deputy spacecraft uses control input to track the target. In this work, we omit the subscript $d$ in the control input, i.e., $u=u_{d}$. The translational equations of motion for spacecraft are given as,
\begin{equation}
\ddot{\vec{r}}_{i} = -\frac{\mu}{r^3_{i}}\vec{r}_{i} + \vec{u}_{i},
\end{equation}
where \(\mu\) stands for the gravitational parameter, \(\vec{r}_{i}\) is the spacecraft position vector and \(r= \| \vec{r} \| \) is its 2-norm.

\subsection{Discrete Time Linear Quadratic Controller}\label{subsec2}
Since the TSG is an add-on scheme, we need a nominal controller that is (locally) stabilizing to a target along the reference orbit. In this work, the discrete-time linear quadratic (DTLQ) controller is implemented to provide an optimal control solution stabilizing the Deputy spacecraft to the target.
% In this work, we use the discrete-time linear quadratic (DTLQ) controller to stabilize the Deputy spacecraft to a target state along the reference orbit. We chose the DTLQ controller because it integrates seamlessly with the Time Shift Governor, an add-on scheme that complements with the inner loop controller. This ensures that the inner loop controller effectively stabilizes the system.

While the Chief spacecraft is assumed to follow a reference orbit, which is an unforced natural motion, the Deputy spacecraft motion is controlled by the feedback controller 
\begin{equation}
\label{eq:control_input}
    u_{d}(t) = K( t_{k} )(X_d(t_{k}) - X_v(t_{k})), \quad t_{k} \leq t < t_{k+1},
\end{equation}
where $X_{v}(t)$ is the virtual target, which is determined by the TSG, and $K$ is the periodic LQR gain which is calculated for the linearization of \eqref{eq:equation_of_motion} along the state trajectory on the reference orbit. To save computational effort, the standard infinite-horizon LQ control gain $K$ is pre-computed for an orbit as
\begin{equation}
\label{eq:control_LQRgain}
    K(t_{k}) = (B^{\T}_{d} S B_{d} + R)^{-1} B^{\T}_{d} S A_{d},
\end{equation}
where $S$ is the infinite horizon solution of the Discrete Algebraic Riccati Equation (DARE),
\begin{equation}
\label{eq:control_DARE}
S = Q - A_d^{\sf T}(t_{k}) S B_d(R + B_d^{\sf T}(t_{k}) S B_d(t_{k}))^{-1} B_d^{\sf T}(t_{k}) S A_d(t_{k}) + A_d^{\sf T}(t_{k}) S A_d(t_{k}),
\end{equation}
where \( Q \succeq 0 \) and \( R \succ 0 \) are the weighting matrices associated with the state and control, respectively, and \( S \) stands for the solution of DARE at the time instant \( t_k \), e.g., \( S = S(t_{k}) \). Here \( A_d(t_{k}) \) and \( B_d(t_{k}) \) represent the discrete-time linearized model at the virtual target state. Note that we assume that the nominal controller, i.e., the DTLQ controller, is able to track a virtual target \( X_v(t) \), which is a time-shifted state trajectory of the reference orbit. 
% This assumption can be held by the solutions of the TSG that ensure its attractivity to the virtual target in a sufficiently long prediction horizon.

\subsection{Constraints}\label{subsec2}
In this paper, we consider the state and control input constraints for the Deputy spacecraft. During the RD mission, the Deputy spacecraft is maintained within the safe approach corridor from the docking port, assuming that the docking port is towards the opposite direction of the Chief spacecraft's velocity direction. This ensures that the docking port of the Deputy spacecraft and the Chief spacecraft is aligned. The Line of Sight (LoS) constraint is written as 
\begin{equation}
\label{eq:constraint_cone}
h_1 = -\frac{v(X_c)^{\T} p(X_d - X_c)}{\|v(X_c)\| \|p(X_d - X_c)\|} + \cos(\alpha)  \leq 0,
\end{equation}
where \(\alpha\) is a LoS half-cone angle.

The physical limit of the thrust magnitude is expressed as
\begin{equation}
\label{eq:constraint_control_input}
h_2 = \|u_d\| - u_{\tt max} \leq 0,
\end{equation}
where \( u_{\tt max} \) is the maximum total magnitude of the physical thrusters. Considering that the spacecraft typically uses a single main thruster for maneuvers, we assume that the attitude control system accurately aligns the spacecraft consistently with the desired thrust direction. 
% thruster nozzle direction accurately aligns with the desired thrust direction, which means that the attitude control problem is not incorporated. 
To achieve faster response, we enforce the constraint in Eq.~\eqref{eq:constraint_control_input} using a saturation function rather than addressing Eq.~\eqref{eq:constraint_control_input} as a constraint by the TSG. The control input, limited by the saturation, is determined as
\begin{equation}
\label{eq:saturated_control_input}
u_d(t) := \min(\|u_d(t)\|, u_{\text{max}}) \cdot \hat{u}_d(t),
\end{equation}
where $\hat{u}_d(t) = u_d(t)/\|u_d(t)\|$. Note that the prediction of the closed-loop response in the TSG accounts for the saturated control in Eq.~\eqref{eq:saturated_control_input} to enforce Eq.~\eqref{eq:constraint_control_input}.

We limit the relative velocity magnitude of the Deputy spacecraft with respect to the Chief spacecraft to avoid the risk of high relative velocity collisions. Hence, the soft docking constraint is defined as
\begin{equation}
\label{eq:constraint_softdocking}
h_3 = \| v(X_d - X_c) \| - \gamma_2\| p(X_d - X_c) \| - \gamma_3 \leq 0,\; \text{if }\|p(X_d - X_c)\| \leq \gamma_1,
\end{equation}
where \( \gamma_1 \), \( \gamma_2 \) and \( \gamma_3 \) are predetermined mission specific parameters.

\begin{figure}[htb]
\centering
\includegraphics[width=1.0\textwidth]{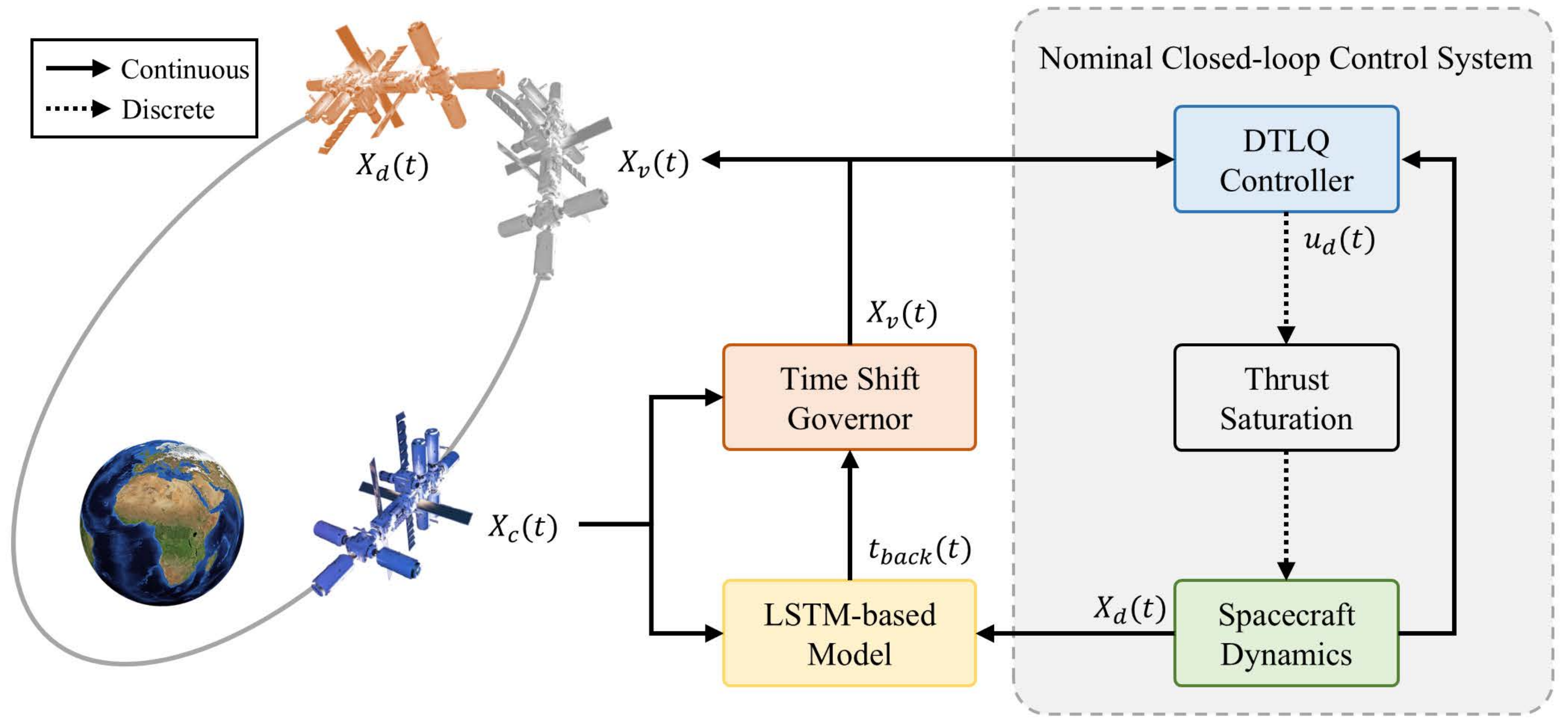}
\caption{Diagram of the control system architecture incorporating learning-based predictive model with TSG and DTLQ controller.}
\label{fig:ControlDiagram}
\end{figure}

\section{Time Shift Governor} \label{sec:time_shift_governor}
As a variant of the parameter governor~\cite{kolmanovsky2006parameter}, the Time Shift Governor (TSG) is an add-on scheme that enforces state and control constraints, see in Fig.~\ref{fig:ControlDiagram}. In this paper, we apply TSG to the spacecraft rendezvous and docking (RD) problem. TSG generates a virtual target trajectory, which is determined as time-shifted trajectory of the Chief spacecraft, by solving a one dimensional constrained optimization problem. 

More specifically, assuming that the Deputy spacecraft is behind the Chief spacecraft in orbital track, to satisfy the constraints, the TSG computes the smallest in magnitude non-positive time shift, which represents the minimum difference in time between a virtual target and the Chief spacecraft along the orbital track, and such that the predicted trajectory assuming the time shift is maintained constant at that value is guaranteed to satisfy the constraints. The TSG then provides the virtual target corresponding to this time shift, i.e.,
\begin{equation} \label{eq:Shift_Virtual}
    X_{v} (t) = X_{c}(t+t_{\tt back}(t)),
\end{equation}
where $t_{\tt back}(t)=t_{\tt back}(t_{k})$ for $t\in [t_{k}, t_{k+1})$ is the time shift, which is thus a piecewise constant function of time. Additionally, the LQ control gain $K$ is precomputed for one orbit period and implemented using linear interpolation. 
% In this work, we assume that the Chief spacecraft is positioned forward along the orbital track relative to the Deputy spacecraft. 
% Note that the upper and lower bounds of the time shift are set to be zero and the initial time shift, respectively.

For TSG to be able to successfully enforce the constraints, at the initial time instant, there must exist a feasible time shift that guarantees that the predicted trajectory corresponding to that time shift will satisfy constraints over a sufficiently long horizon. 
For TSG to be able to gradually converge the time shift to zero (i.e., for Deputy spacecraft to actually be able to reach the Chief spacecraft), TSG must be able to make at least small adjustments in the time shift when the Deputy spacecraft state is sufficiently close to the state on the target orbit corresponding to the constant value of the time shift without causing constraint violations by the predicted trajectory.
% The virtual target determined by the TSG needs the following properties to demonstrate its effectiveness. First, at the initial time instant, there exists a feasible time shift that guarantees a predicted trajectory satisfying constraints over a sufficiently long horizon. The nominal controller is then capable of (locally) stabilizing the Deputy spacecraft to a target. Lastly, with any time shift parameter $t_{\tt back}$, the steady-state response strictly satisfies the constraints, which means that the Deputy spacecraft with TSG satisfies the constraints for all future time. 

In the implementation of TSG in this paper, we use the prediction horizon of one orbital period of the Chief spacecraft to check whether the constraints are satisfied. The initial time shift parameter is determined by finding the state along the nominal trajectory of the Chief spacecraft closest to the state of the Deputy spacecraft. At the subsequent time instants, bisections are used to solve the one dimensional optimization problem and compute the time shift where the search interval is restricted to the one between the previous time shift (feasible by construction) and zero.
% Considering the same relative dynamics repeated every orbit period, we use one orbit period of prediction horizon in a moving horizon manner. The initial time shift parameter is computed using the brute force algorithm. For computational efficiency, the search order begins at the point along the reference state trajectory closest to the Deputy spacecraft state. In other words, we adopt a warm start approach.

\section{Learning-based Time Shift Governor (L-TSG)} \label{sec:deep_learning_tsg}
% In this section, we introduce a novel framework that integrates the Time Shift Governor (TSG) with Long Short-Term Memory (LSTM) networks. The LSTM-based TSG provides a time shift with lower computational effort while enforcing constraints. We apply a phase-adaptive sliding window approach to leverage the computational efficiency of the proposed deep learning scheme during the entire RD mission. During training stages, a customized loss function is used to avoid overconfidence in the solutions of our prediction model. Our proposed deep learning scheme is then applied in a hybrid prediction manner to estimate the time shift parameter. 

In this section, we introduce a novel framework that accelerates the Time Shift Governor (TSG) onboard computations through the use of a deep-learning model to approximate the solution of the TSG optimization problem thereby providing an explicit or an imitation learning based solution. This deep learning model utilizes a Long short-term memory (LSTM) layer and a fully connected layer to map sequential data input into the time shift parameter. The proposed L-TSG computes the time shift parameter which results in enforcing constraints and accomplishing the rendezvous mission. We apply a phase-adaptive sliding window (PA-SW) approach to enhance the performance of the deep learning model during the RD mission. It is important to note that, in an ideal case, the deep learning prediction model can generate the time shift parameter just as a function of the combined state of the Deputy and Chief spacecraft at a given time instant.
% corresponding to a state at a given time instant since it is proposed as a mapping function. 
However, the deep learning model may be able to compute a more accurate approximation of the optimal time shift from a sequence of past state instead of only the current state as it may be able to more easily evolve necessary representation during training.
% time shift prediction by providing additional input data, such as a sequence of past states, compared to the time shift prediction using only the current state. 
In the training process, we utilize a loss function based on a problem-specific heuristic. 
% Thus the proposed model provides an imitation learning based or an explicit TSG implementation. 

The time shift prediction from the LSTM model has an approximation error, which is a potential reason for causing constraint violations. L-TSG first checks constraint violation over the prediction horizon and then determines whether it updates the time shift. If the time shift prediction is valid, the L-TSG uses the time shift while, if not, applying the conventional TSG with a set of time shifts, i.e., $\mathbb{R}_{[t_{\tt back}^- 0]}$, which is bounded by the previous time shift parameter $t_{\tt back}^-$ and zero.  With this framework, L-TSG reduces an average computation time and bounds the worst-case computation time to that of the conventional TSG.

% {\color{blue}This model is then integrated with the TSG, which uses the bisection method to handle cases where the output of the deep learning model does not enforce constraints due to approximation errors. In such cases, the bisection search is initialized with a lower bound set to the last valid time shift parameter from the previous step and an upper bound fixed at zero. This approach ensures that the search interval includes a feasible solution while progressively reducing the time shift to zero. If no valid lower bound is available (e.g., at the beginning of the mission), the bisection search starts with a predefined range determined by mission parameters to guarantee feasibility. }
% {\color{red} Question for you: How is this done?  For bisections, one needs a lower bound and an upper bound. So if the neural network generated value does not satisfy the constraints, how are lower bound and upper bound determined?  or do you do bisections search from scratch at that point.}

\subsection{Data Preparation}\label{subsec2}
The dataset consists of 500 trajectories corresponding to different initial states of the Deputy spacecraft, $X_{d}(t_0)$, and with the conventional TSG based on bisections implemented to enforce the constraints. We allocate the dataset as follows: 60\% for training, 20\% for validation, and 20\% for testing. To enhance the training performance, we use Min-Max normalization to scale the time shift values to a range between zero and one, i.e., $t^{(\tt min/max)}_{\tt back}\in [0,1]$, maintaining the original distribution of the data without distortion of information. This facilitates faster convergence of the neural network training.
\subsection{Network Architecture}\label{subsec2}
\begin{figure}[htb]
\centering
\includegraphics[width=0.7\textwidth]{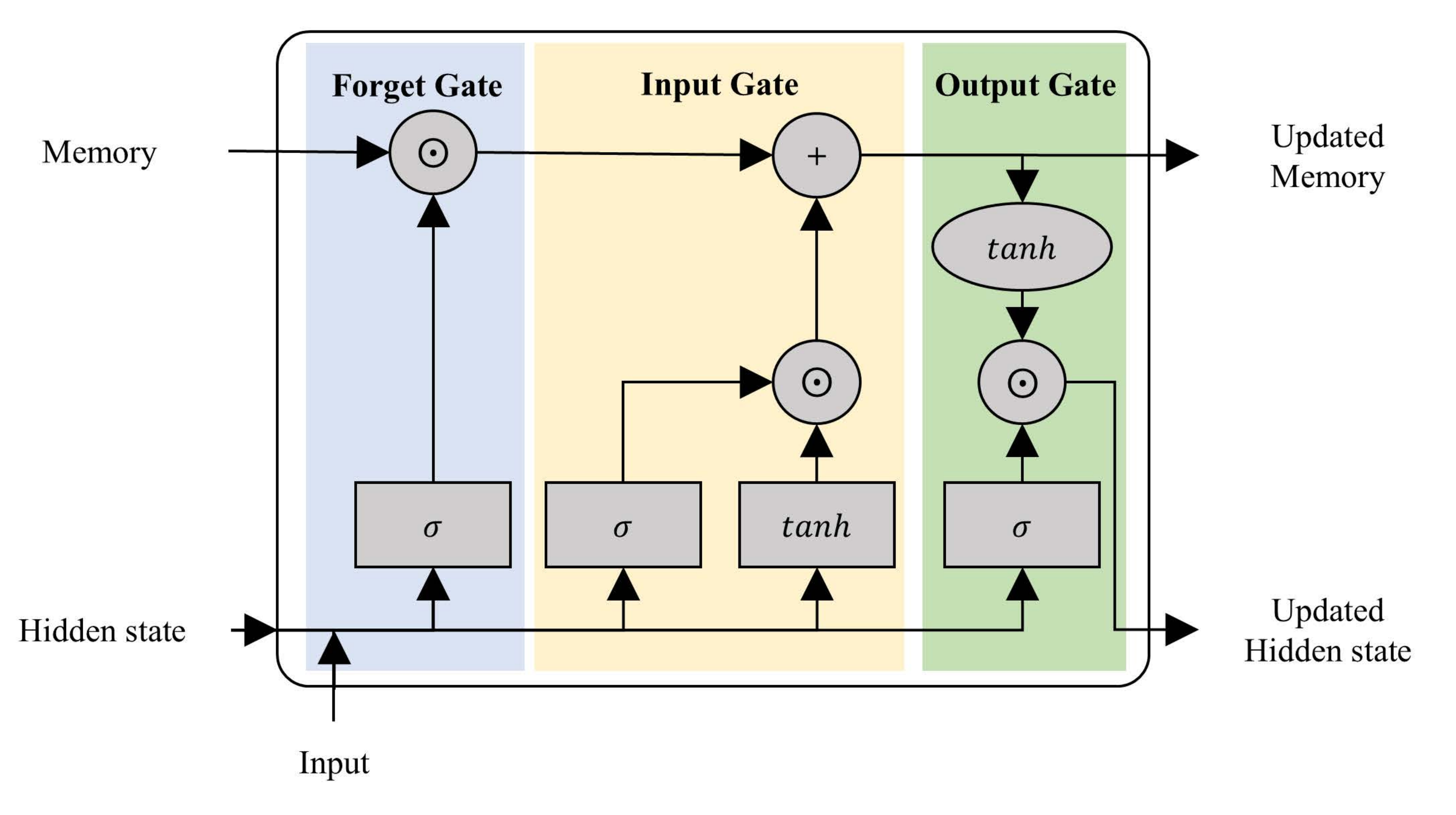}
\caption{Diagram of the LSTM cell, illustrating the input, forget, and output gates.}
\label{fig:lstm_cell}
\end{figure}

In this study, we use the LSTM network with the input, output, and forget gates in Fig.~\ref{fig:lstm_cell} for the time shift parameter prediction model. LSTMs are often used to capture temporal dependencies in sequential data by maintaining a memory of previous states through their gating mechanism~\cite{han2019review, lara2021experimental}. 
% The input, output, and forget gates in Fig.~\ref{fig:lstm_cell} process the flow of information, selectively retaining the spacecraft motion with respect to safety constraints over long sequences. 
The cell state updates utilize the Hadamard product, $\odot$, allowing for element-wise multiplication of the input, forget, and cell gates. The proposed LSTM network maps a sequence of the Chief and Deputy spacecraft's states to the time shift parameter.

The use of sequential data allows LSTM network to account for temporal dependencies, that could be important for accurate predictions. The LSTM cell encodes the sequential information of the spacecraft states, encapsulating the resulting trajectory and imposed constraints. The fully connected neural network then decodes the encoded data from the state sequence to the output of the predicted time shift. Compared to the Multilayer Perceptron (MLP), the LSTM cell can achieve better performance by capturing temporal dependencies in the sequential data~\cite{ahmed2022review}.

The LSTM-based model is thus a mapping from the past sequence of Chief and Deputy spacecraft states to the time shift parameter:
% , $\Gamma: \R^{12}\times \N^{\tt card(\mathcal{I}_{w})}\to \R_{\leq 0}$, which maps from sequential state space to the time shift space, is defined as
\begin{equation}
{t}_{\tt back} = \Gamma_{\tt Pred} (W(t_k)),
\end{equation}
where $W(t_k)$ is a window of sequential data at time step $t_k$.

\begin{figure}[htb]
\centering
\includegraphics[width=0.9\textwidth]{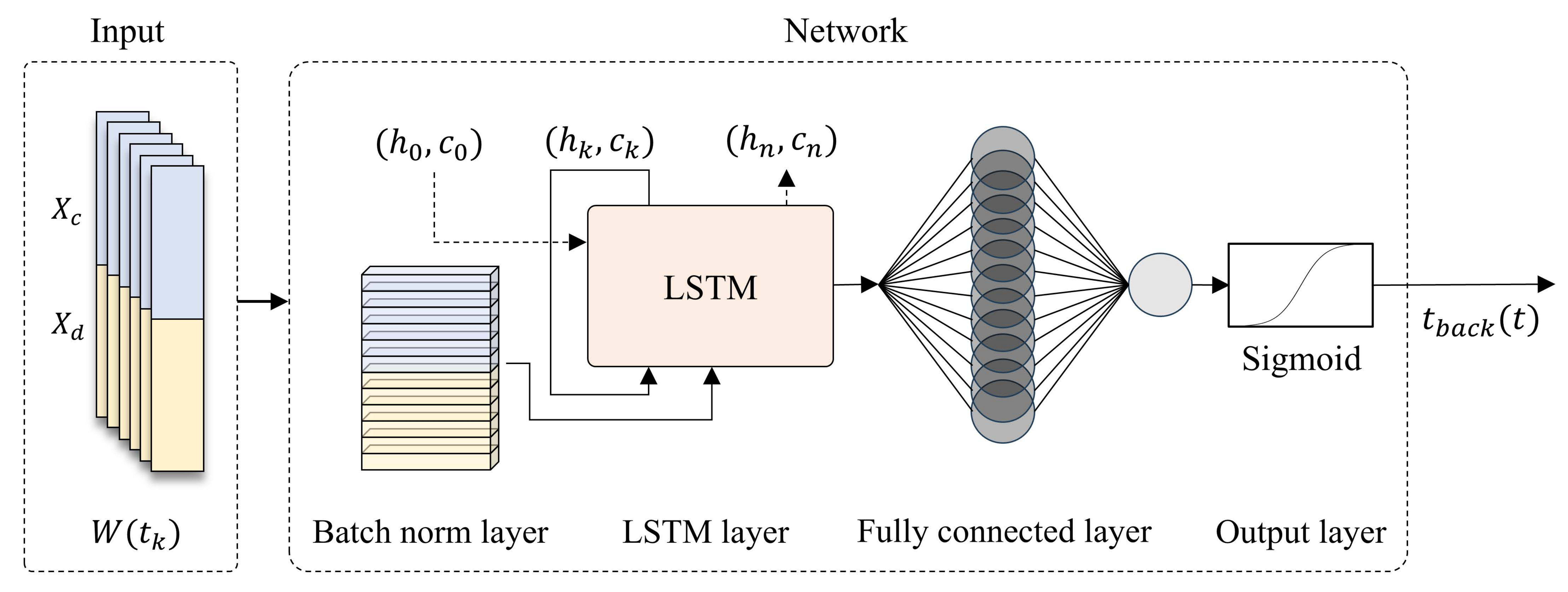}
\caption{Overview of LSTM-based time shift prediction model network architecture.}
\label{fig:ModelArchitecture}
\end{figure}

Fig.~\ref{fig:ModelArchitecture} illustrates the network architecture, implemented using PyTorch. To enhance stability and efficiency in training, we first apply a batch normalization layer to standardize inputs. An LSTM layer then processes the sequential data in a specified window. To address the overfitting problem, we use the dropout layer to randomly disable a portion of the inputs. Then, a fully connected layer translates the hidden states of the LSTMs as inputs for the sigmoid function. The sigmoid function constrains the output values between 0 and 1, aligning with the range established by the Min-Max scaling method applied during data preparation, ensuring consistency in the network's predictions. 
% Moreover, the sigmoid function's smooth gradient helps our neural network perform a stable learning process during training.

\subsection{Phase-Adaptive Sliding Window (PA-SW) Approach}\label{subsec2}
% The DL model makes decisions based on a finite sequence of past deputy and chief spacecraft states, within a specified past window of data. 
We implement a PA-SW approach to enhance the efficiency and accuracy of our prediction model during the RD mission. This approach involves dynamically adjusting the window size based on the current phase of the mission, defined by the relative distance between the Deputy and Chief spacecraft. Sliding window approaches have been previously used to handle sequential data~\cite{yu2014time, hota2017time}, and we modify this approach to implement an effective prediction model corresponding to the specific phases of the RD mission.
By adapting the window size according to the mission phase, our approach ensures efficient data utilization and computational cost reduction throughout the mission.

For the Crew-3 mission, which includes two phases (distances over 1km, and 1km to 0km), three models are used distinguished by window sizes: window size of $w_k = 1$ for the initial time, $w_k = 2$ for the first phase (over 1km), and $w_k = 3$ for the second phase (1km to 0km). This ensures that as the spacecraft gets closer, the window size increases, enhancing prediction accuracy during the critical close-range phase. For the Molniya orbit mission, which also consists of two phases (distances over 1km, and 1km to 0km), two models are used, both with the same window size of $w_k = 100$. Since a larger window size is already in place, it remains consistent for both phases, ensuring accuracy without requiring further adjustment. The window is adjusted as:
\begin{equation}
W(t_k) = \{ [X^{\T}_{c}(t_{j}), X^{\T}_{d}(t_{j})]^{^{\T}}\in \R^{12 \times w_{k}}\; |\; j\in \mathcal{I}_{w}(t_{k}) \} ,
\end{equation}
where $\mathcal{I}_{w}(t_{k}) = \{ j\in \N \mid j = \max(0, k - w_k + 1), \ldots, k\}$ is a set of indices in the dynamic sliding window at time step $k$ and $w_k$ is the window size, which is set dynamically based on the mission phase.

% For distances between 50km and 1km, the model utilizes small window sizes (1 and 2). As the Deputy spacecraft gets close to the Chief, from 1km to docking, the window size is increased to 3. This larger window increases model complexity and fidelity and enhances the prediction accuracy at critical close-range phases.

% The window is adjusted as
% \begin{equation}
% W(t_k) = \{ [X^{\T}_{c}(t_{j}), X^{\T}_{d}(t_{j})]^{^{\T}}\in \R^{12 \times w_{k}}\; |\; j\in \mathcal{I}_{w}(t_{k}) \} ,
% \end{equation}
% where $\mathcal{I}_{w}(t_{k}) = \{ j\in \N \mid j = \max(0, k - w_k + 1), \ldots, k\} \}$ is a set of indices in the dynamic sliding window at time step $k$ and $w_k$ is the window size, which is set dynamically based on the mission phase: $w_k = 1$ at the initial time, $w_k = 2$ during the first phase (50km to 1km), and $w_k = 3$ throughout the second phase (1km to 0km).

This phase-adaptive strategy ensures that immediate and appropriate predictions are made with the available data, allowing the model to adapt as more data becomes available and as the Chief spacecraft is approached where the constraints become more prominent.

\subsection{Custom Loss Function}\label{subsec2}
We design a loss function that incorporates the mean squared error (MSE) and a heuristic term penalizing predicting small in magnitude time shifts than in the data which could be unsafe. MSE minimizes the average squared discrepancy of predictions with respect to the target values. Specifically, the loss function based on the MSE alone enables a prediction model to generate smaller time shift estimates than in the data which could be unsafe and lead to constraint violation. Thus, beside the MSE in the loss function, we augment a penalty term, which is the mean squared ReLU (MSReLU), with a penalty weight $\eta$, i.e.,

\begin{equation} \label{eq:customizedLoss}
\mathcal{L}_{\tt{Total}} = \mathcal{L}_{\tt{MSE}} + \eta \cdot \mathcal{L}_{\tt{MSReLU},} 
\end{equation}
where
\begin{equation}
\mathcal{L}_{\tt{MSE}} = \frac{1}{N} \sum_{i=1}^{N} (\hat{y}_i - y_i)^2,\quad \mathcal{L}_{\tt{MSReLU}} = \frac{1}{N}\sum_{i=1}^{N}({\tt{ReLU}}(\hat{y}_i - y_i))^2,
\end{equation}
and where \(\hat{y}_i\) and \(y_i\) denote the predicted time shift and the target time shift, respectively, and \(N\) is the total number of samples. MSReLU applies penalties when prediction errors are positive, considering that the target time shift is the maximum time shift to satisfy the constraints. In the other setting with the Deputy spacecraft approaching from the positive V-bar direction, the MSReLU function becomes $\mathcal{L}_{\tt MSReLU}=\frac{1}{N}\sum_{i=1}^{N}({\tt ReLU}(y_{i}-\hat{y}_{i}))^{2}$. Note that the use of the square in MSReLU term enhances differentiability and improves unit matching with the MSE term. 
% The integration of the MSReLU term guides our model to enforce the constraints while maintaining optimality.

\subsection{Hyperparameter Tuning}\label{subsec2}
In optimizing hyperparameters for the LSTM-based prediction model, we use the grid search algorithm to find the optimal hyperparameter combination within the accessible hyperparameter space. The hyperparameters are defined as follows: A hidden state space $ \{ h \in \N \mid 2^{s},\; s=6,7,8 \}$, dropout rate $\delta \in \{ 0.1,0.15,\cdots,0.3\}$, batch size space $\{ N \in \N \mid 2^{s},\; s=7,8,9,10,11 \}$, learning rate space  $\{\lambda \in \R \mid s_1 \times 10^{s_2}, s_1=5, 2, 1, s_2=-3,-4,-5\}$, and the penalty weight $\eta \in \{0.1, 0.5, 1.0, 10, 100\}$ in Eq.~\eqref{eq:customizedLoss}. We utilize a maximum of 300 epochs and apply an early stopping mechanism with a patience parameter of 15 epochs to save training time and prevent overfitting.

% We adapt the learning rate based on the phase-specific window size: a learning rate of $0.001$ is used for the first phase (window sizes 1 and 2), and a reduced rate of $0.0002$ for the second phase (window size 3) to enhance precision and stability as the spacecraft approaches docking.

Moreover, to enhance the robustness and generalizability of our model, we incorporate a 10-fold cross-validation process. This method splits the dataset into 10 subsets, using each part once as the validation set while the remaining 9 parts are used as a training set. 
% This approach ensures that the model is tested across a diverse range of scenarios. 
The optimal hyperparameters for each window size were those that minimized the mean validation loss of the cross-validation process, which was set to be the same custom loss function used during training. This consistency in loss functions across training and validation stages ensures that the performance metrics are directly comparable, enhancing the reliability of our model evaluation. 

\subsection{Hybrid TSG Algorithm}\label{sec:IterativePrediction}
% Given the inherent uncertainties associated with deep learning predictions, we extend our methodology by integrating a bisection algorithm as a fallback mechanism within our prediction algorithm. This hybrid approach ensures optimal decision-making by complementing the adaptability and speed of deep learning with the reliability and precision of deterministic numerical methods. Employing this strategy allows fine-tuning the time shift parameter, \(t_{\text{back}}\), under varied operational scenarios, thereby enhancing the robustness and accuracy of our system. The hybrid prediction algorithm plays a critical role in guaranteeing the reliability of $t_{\tt{back}}$ estimates computed from our prediction model.
We combine a learning-based model with a TSG that uses the bisection algorithm. The learning-based model computes the time shift as a function of the states which is then verified through forward simulations; if the verification fails, the algorithm reverts to the bisection-based TSG. This approach reduces the computing time of the time shift parameter in most of the cases. More specifically, 
% with an algorithm designed to validate and apply the predicted time shifts in a manner analogous to control input application in MPC. The algorithm’s similarity to MPC lies in its predictive and recursive adjustment of the control action---here, the time shift parameter $t_{\text{back}}$---to optimize for constraint satisfaction and system performance over a prediction horizon. 
our proposed hybrid algorithm computes a time shift \( t_{\tt{back}} \) based on \(W(t_{k}) \) as follows:
\begin{enumerate}
    \item \textbf{Time Shift Estimation}:
    Every update period \( P_{\tt TSG} \), our learning-based model estimates the time shift \( t_{\tt back}\) corresponding to \(W(t_{k}) \), e.g., \(t_{\tt back}=\Gamma_{\tt Pred}(W(t_{k})) \).

    \item \textbf{Virtual Target State Computation}:
    The state of the virtual target, \( X_v(X_{c}(t), t_{\tt back}) \), is updated to the state of the Chief spacecraft corresponding to the estimated time shift in Eq. (\ref{eq:Shift_Virtual}).

    \item \textbf{Estimated Time Shift Verification}: 
    We check constraint satisfaction over the predicted trajectories of the Deputy and Chief spacecraft, for a sufficiently long prediction horizon \(T_{\tt TSG} \), with the virtual target corresponding to the time shift estimated by the learning-based model, based on Eqs. (\ref{eq:equation_of_motion}) and (\ref{eq:control_input}). If the estimated time shift parameter is verified, the virtual target is updated corresponding to the time shift; otherwise, the previous virtual target holds.

    \item \textbf{Verification Failure Handling}:
    Conventional TSG based on bisections is applied to determine the time shift.
\end{enumerate}

% This process, summarized in Algorithm~\ref{alg:iterative_prediction}, where \(P_{\tt TSG}\) is the horizon for trajectory prediction, not only ensures the validation of estimated time shifts against the system's constraints but also acts as a safety filter for enhanced model dependability. By addressing the risks inherent in the prediction model, this algorithm enhances the reliability of trajectory planning for a successful mission.

% The verification process for the estimated time shift involves checking constraint satisfaction over the prediction horizon. During this step, the system computes inequality constraints \(h_1(X_{c},X_{d})\) and \(h_2(X_{c} (\tau+t_{\tt back}),X_{d},u_{d}(\cdot))\). If the relative distance \( \|p(X_{d} - X_{c})\| \) is less than or equal to \( \gamma_1 \), an additional constraint \(h_3(X_{c},X_{d})\) is computed. If any of these constraints are violated, the validation is marked as unsuccessful (\(\mathds{1}_{\tt{safe}} = {\tt False}\)). If the validation is successful (\(\mathds{1}_{\tt{safe}} = {\tt True}\)), the algorithm proceeds with the time shift. The validation is repeated for each time step \( \tau \) in the prediction horizon, with forward propagation of the Chief, Deputy, and virtual target states for one prediction step \( \Delta \tau \).

These steps are summarized in Algorithm~\ref{alg:iterative_prediction}. We use $\epsilon = 1 \times 10^{-10}$ in the implementation.

\begin{algorithm}
\caption{Hybrid TSG Algorithm }
\label{alg:iterative_prediction}
\begin{algorithmic}[b]
% {\color{blue}
\State \bf{Input: } $ t_n, W(t_n), t_{\tt{back}}(t_{n-1}) $
\State \bf{Output: } $ t_{\tt{back}}(t_n) $
\State $k_{1}$, $k_{2}$ = 0, 0;
\For{$k$ in \(\{1,2,3,\ldots, N_{\tt sim}\}\) }
    \State Estimate a time shift parameter using the learning-based model: $ t_{\tt{back}} = \Gamma_{\tt Pred}(W(t_{k}))$;    
    \State \(\mathds{1}_{\tt{safe}}= \text{Verification}(t_k,X_{c},X_{d},t_{\tt back}, T_{\tt TSG})\);
    
    \If{$ \mathds{1}_{\text{safe}} $}
        \If{$ |t_{\tt back}(t_k)-t_{\tt back}(t_{k-1})|<\epsilon$}
            \State $k_{1}=k_{1}+1;$
            \If{$ k_{1} < N_{1}$}
                \State Use $ t_{\tt back}(t_k) $ computed from the learning-based model.
            \Else
                \State Perform bisection search with bounds: $ t_{\text{back}} \in [t_{\text{back}}(t_{k-1}), 0] $
                % \State Compute $ t_{\text{back}}(t_n) $ using the bisection algorithm.
                \State $k_{1}=0$;
            \EndIf
        \EndIf        
    \ElsIf{$ \mathds{1}_{\text{safe}} = {\tt False} $}
        \If{$k_{2}<N_{2}$}
            \State $t_{\tt back}(t_{k})=t_{\tt back}(t_{k-1})$;
            \State $k_{2}=k_{2}+1$;
        \Else
            \State Perform bisection search with bounds: $ t_{\text{back}} \in [t_{\text{back}}(t_{k-1}), 0] $
            % \State Compute $ t_{\text{back}}(t_n) $ using the bisection algorithm. 
            % \State {\color{red}how? see my note above. Also the implication could be that the worst-case computing effort will be worse than bisections alone alone; hopefully this is relatively rate}
            \State $k_{2}=0$;        
        \EndIf        
    \EndIf
\EndFor
\end{algorithmic}
\end{algorithm}

\section{Simulation Results} \label{sec:results}
In this paper, we demonstrate the learning-based TSG (L-TSG) for spacecraft rendezvous and docking missions in circular and elliptic orbits which correspond to: Low Earth Orbit (LEO) used in the Crew-3 mission and the Molniya orbit. 
% Each scenario highlights the capability of L-TSG in different orbital environments.
% In this paper, we demonstrate the deep learning-accelerated TSG (DL-TSG) via the Crew-3 mission scenario. We design a scenario with an initial distance of about 50 km between two spacecraft: the Chief spacecraft orbits along the orbit of the International Space Station (ISS), and the Deputy spacecraft starts 50 km behind the Chief spacecraft. 

% Orbital element
\begin{table}[b]
\centering
\caption{Classical orbital elements of the reference ISS orbit.}
\label{tab:orbital_elements}
\begin{tabular}{lccccc}
\hline
SMA, \( a \) & Eccentricity, \( e \) & Inclination, \( i \) & RAAN, \( \Omega \) & Argp, \( \omega \) \\
\hline
6798.281637 [km] & 0.000551 & 0.900516 [rad] & 5.909781 [rad] & 1.872335 [rad] \\
\hline
\end{tabular}
\end{table}

\begin{table}[b]
\centering
\caption{Orbital elements of the reference Molniya orbit.}
\label{tab:orbital_elements_molniya}
\begin{tabular}{lccccc}
\hline
SMA, \( a \) & Eccentricity, \( e \) & Inclination, \( i \) & RAAN, \( \Omega \) & Argp, \( \omega \) \\
\hline
26646.680769 [km] & 0.74 & 1.096067 [rad] & 0.0 [rad] & 4.88692 [rad] \\
\hline
\end{tabular}
\end{table}

\subsection{Simulation Specifications}\label{subsec2}
In this simulation, we consider the application of our learning-based TSG (L-TSG) to Crew-3 RD mission and highly elliptic orbit, the Molniya orbit. We consider the Chief spacecraft following the reference orbit, where the orbital elements are specified in Tables \ref{tab:orbital_elements} and \ref{tab:orbital_elements_molniya}. The LEO and Molniya orbit have a period of 92.97 minutes and 721.48 minutes, respectively. The one orbital period \( P_{\text{ref}} \) is chosen as the prediction horizon for the TSG at a time instant \( \tau \), i.e., the prediction is performed over the time interval, \( [\tau, \tau + P_{\text{ref}}] \). The gravitational parameter is \( \mu = 398600.4418\text{km}^3/\text{sec}^2 \).

The nominal controller of the Deputy spacecraft is the DTLQ controller in Eq. (\ref{eq:control_input}) with an LQR gain \( K(t_k) \) computed for the following state and control weighting matrices, $Q = \text{diag}(10, 10, 10, 1, 1, 1), \; R = \text{diag}(1, 1, 1).$ Since LTSG locates the virtual target state along the reference trajectories, the LQR gain is precomputed over one orbital period before simulations begin. Using linear interpolation, the nominal controller applies the precomputed LQR gain, which is a piecewise constant as a function of time.

% The LQR gain \( K(t_k) \) is computed from a linearized model at the virtual target, corresponding to \( t_{\tt{back}} \).
% {\color{blue}
% The LQR gain \( K(t_k) \) is precomputed for a range of \( t_{\tt{back}} \) values using the linearized model at the virtual target. 

% option1: During the TSG adjustment process, if \( t_{\tt{back}} \) does not match a precomputed value, \( K(t_k) \) is interpolated, avoiding repeated solutions of the Riccati equation and reducing the computational cost.

% option2: If the required \( t_{\tt{back}} \) is between precomputed points, \( K(t_k) \) is interpolated, avoiding repeated calculations and reducing the computational cost.
% }
% {\color{red}Question$: $This means that Riccati equation has to be solved multiple times as $t_back $is adjusted by bisection TSG?  Before $(5)$, it says the gain is precomputed.  How can you precompute it if you do not know what $t_{back}$ is???} 

% Contraint spec
In our simulations, the learning-based TSG manages the constraints specified in Eqs. (\ref{eq:constraint_cone}), (\ref{eq:constraint_control_input}), and (\ref{eq:constraint_softdocking}) with parameters as follows: The half-cone angle \( \alpha \) is set to \( 20^\circ \), the thrust magnitude limit \( u_{\text{max}} \) to \( 0.5 \) m\(\cdot\) s\(^{-2}\), and the approach velocity constraints are set with \( \gamma_1 = 5 \) km for distance, \( \gamma_2 = 20 \) rad \(\cdot\) s\(^{-1}\) for angular rate, and \( \gamma_3 = 0.001 \) km\(\cdot\) s\(^{-1}\) for relative velocity limit. The Crew Dragon spacecraft, which we utilize in both missions, has a launch mass of 12,519 kg~\cite{Heiney2020} and is equipped with 16 Draco engines, each capable of a maximum thrust of 400N~\cite{SpaceXDragon}, resulting in an actual thrust magnitude of 0.5112 m\(\cdot\) s\(^{-2}\). To ensure safety, we have conservatively set the thrust magnitude limit to 0.5m\(\cdot\) s\(^{-2}\).

% To further enhance the realism and robustness of our simulation of the Crew-3 RD mission, we incorporate perturbations into the initial state of the Deputy spacecraft. Specifically, we introduce Gaussian noise where the covariance matrix, \( \Sigma \), is scaled proportional to the differences between the initial positions and velocities of the Deputy and Chief spacecraft. The elements of \( \Sigma \) are defined such that the standard deviations of the perturbations are \( \frac{1}{10} \) to \( \frac{1}{100} \) of the initial state values, represented as \( \sigma_i = \alpha \times |\Delta X_{\text{rel}_{(d,c)}}| \) where \( \alpha \) ranges from \( 0.01 \) to \( 0.1 \) and \( \Delta X_{\text{rel}_{(d,c)}} \) are the differences in the corresponding initial states. Importantly, among the perturbed initial conditions, we utilized those that satisfied all operational constraints and possessed a valid time shift parameter at the initial step, thereby confirming the feasibility of trajectory planning. This method of introducing perturbations is crucial for validating the DL-TSG's capability to maintain safety and success.

To demonstrate the robustness of the L-TSG, perturbations \(\delta X\) are added to the nominal initial state of the Deputy spacecraft $\bar{X}_{d}(t_{0})$, e.g., \( X_{d}(t_{0})=\bar{X}_{d}(t_{0})+\delta X\). The perturbations on position and velocity are chosen as, respectively, $p(\delta X) \sim \mathcal{N}(0,\sigma^{2}_{\tt pos}\cdot[I]_{3}) ,\; v(\delta X) \sim \mathcal{N}(0,\sigma^{2}_{\tt vel}\cdot[I]_{3}),$ where standard deviations of position \(\sigma_{\tt pos}\) and of velocity \(\sigma_{\tt vel}\) are set to be a tenth of the initial relative distance of the Deputy spacecraft with respect to the Chief spacecraft and a hundredth of the initial relative velocity, respectively, e.g., \(\sigma_{\tt pos}= (0.1)\cdot\| p(\bar{X}_{d}(t_{0}) - X_{c}(t_{0}) )\| \) and \(\sigma_{\tt vel}= (0.01)\cdot\| v(\bar{X}_{d}(t_{0}) - X_{c}(t_{0}) )\| \). 
From these randomly generated initial states we then only retain initial perturbed Deputy spacecraft states \(X_{d}(t_{0})\) that are feasible, i.e., satisfy two criteria:
\begin{itemize}
    \item The states satisfy all the imposed constraints.
    \item The states have an initial time shift parameter that avoids constraint violation over the predicted trajectory for the prediction horizon $P_{\tt ref}$.
\end{itemize}

% Initial condition
\begin{table}[t]
\caption{Expected initial state of the Deputy relative to the Chief, e.g., \( \mathbb{E}\big[X_d(t_{0}) - X_c(t_{0})\big] \).}
\label{tab:initial_condition}
\centering
\begin{tabular}{cccccc}
\hline
\( \delta x_0 \) [km] & \( \delta y_0 \) [km] & \( \delta z_0 \) [km] & \( \delta{\dot x_0} \) [km/s] & \( \delta{\dot y_0} \) [km/s] & \( \delta{\dot z_0} \) [km/s] \\
\hline
-25.9809 & 27.8498 & 22.7715 & -0.0350 & -0.0066 & -0.0234 \\
\hline
\end{tabular}
\end{table}

\begin{figure}[t]
\centering
\includegraphics[width=0.8\linewidth]{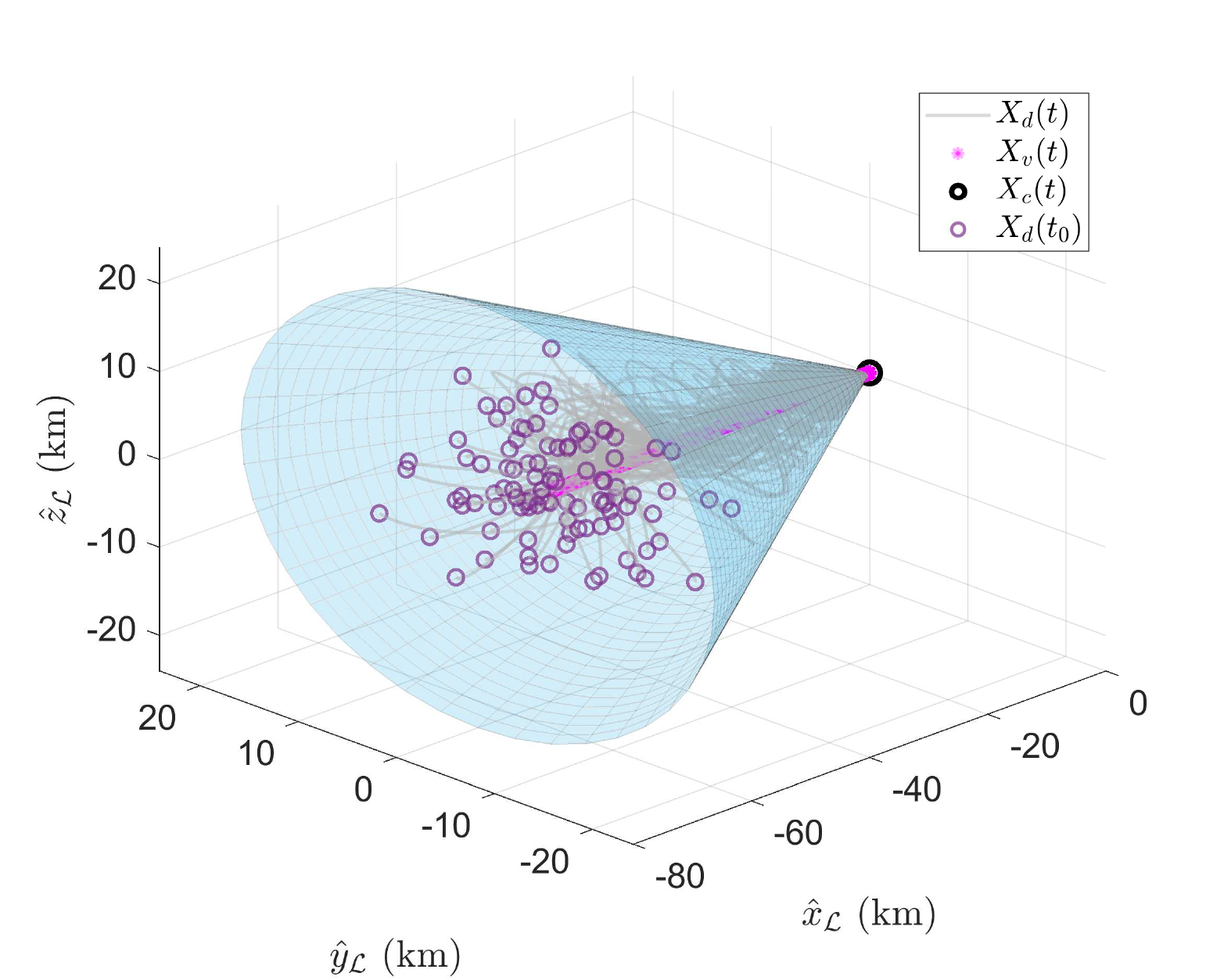}
\caption{Trajectories expressed in the VNB frame: the Deputy spacecraft’s path (gray line) tracks the virtual target’s trajectory (magenta asterisk).}
\label{fig:trajectory_3d_MC}
\end{figure}

\floatsetup[figure]{style=plain}
\captionsetup[subfigure]{labelfont=bf,textfont=normalfont}
\begin{figure}
  \sidesubfloat[]{\label{fig:relposStateChief}\includegraphics[width = 2.5 in]{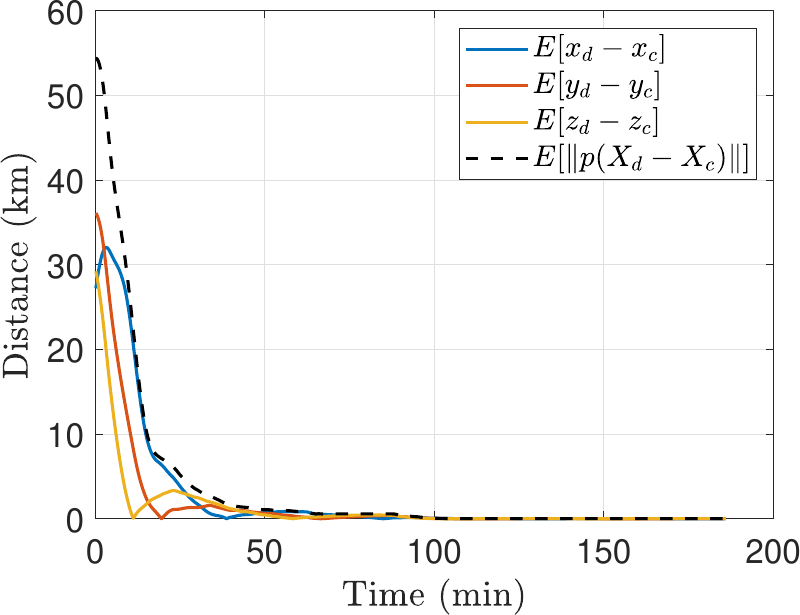}}\;%
  \sidesubfloat[]{\label{fig:relvelStateChief}\includegraphics[width = 2.5 in]{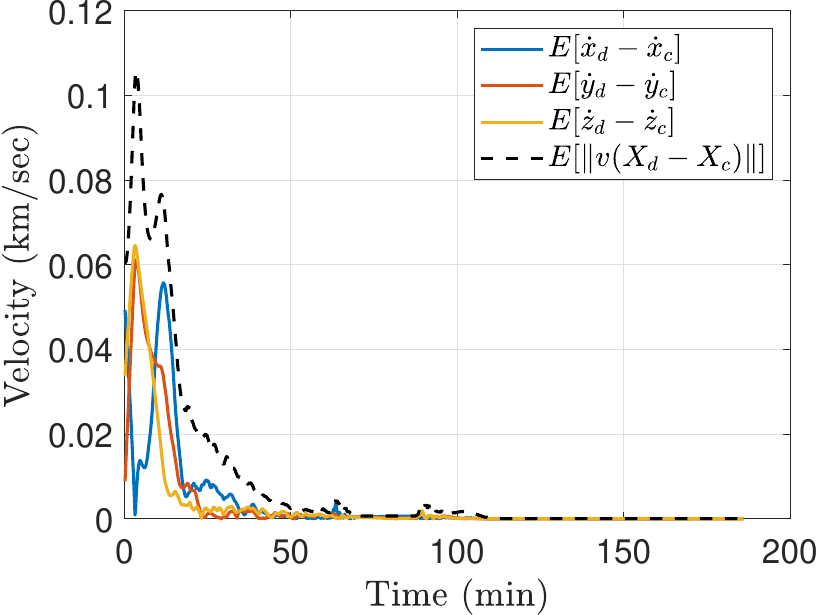}}\\
  \sidesubfloat[]{\label{fig:relposStateTarget}\includegraphics[width = 2.5 in]{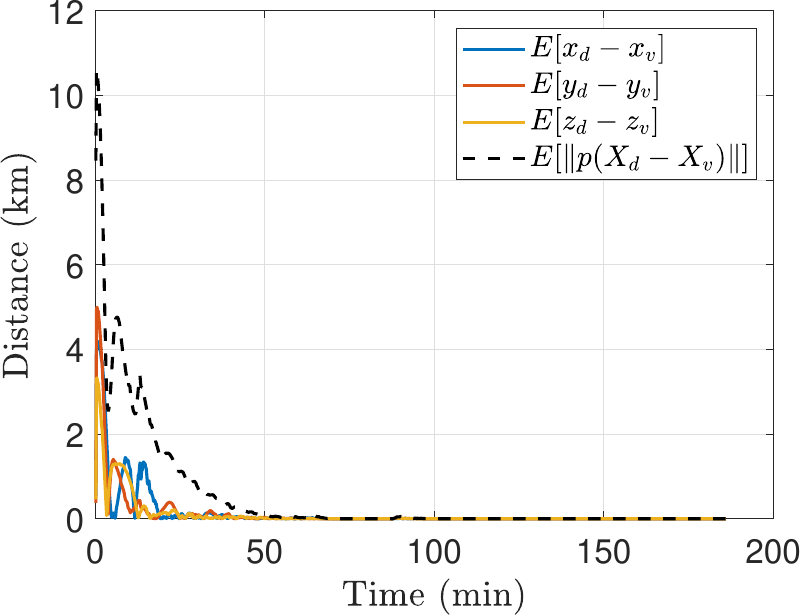}}\;
  \sidesubfloat[]{\label{fig:relvelStateTarget}\includegraphics[width = 2.5 in]{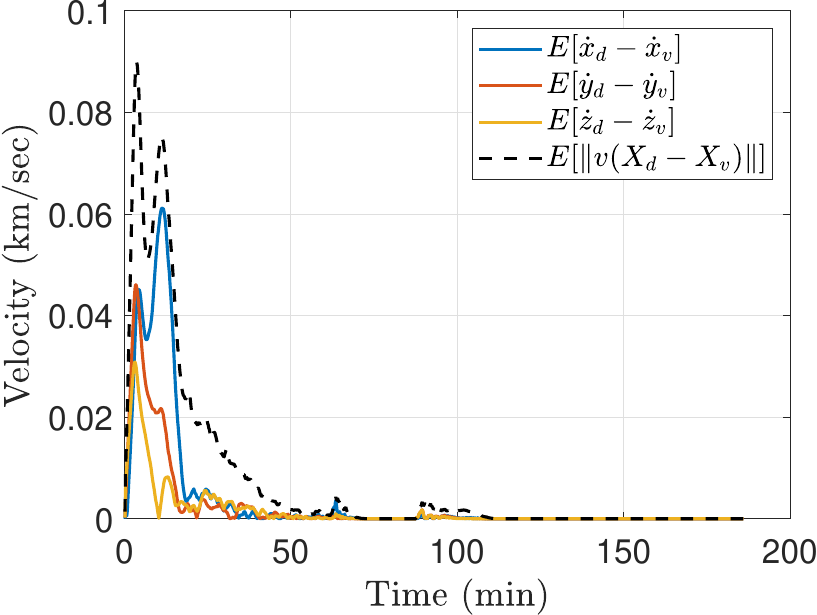}}\;
  \caption{Time histories of relative a) position; b) velocity to the Chief spacecraft; c) position; d) velocity to the virtual target.}\label{fig:rel_state_MC}
\end{figure}

\floatsetup[figure]{style=plain}
\captionsetup[subfigure]{labelfont=bf,textfont=normalfont}
\begin{figure}
  \sidesubfloat[]{\label{fig:constraint_h1}\includegraphics[width = 1.85 in]{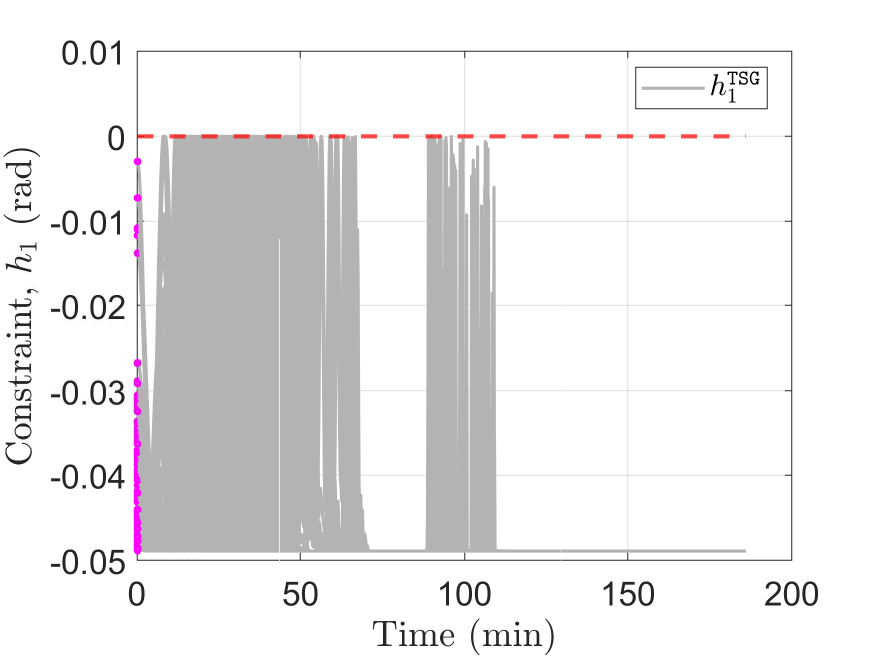}}\;%
  \sidesubfloat[]{\label{fig:constraint_h2}\includegraphics[width = 1.85 in]{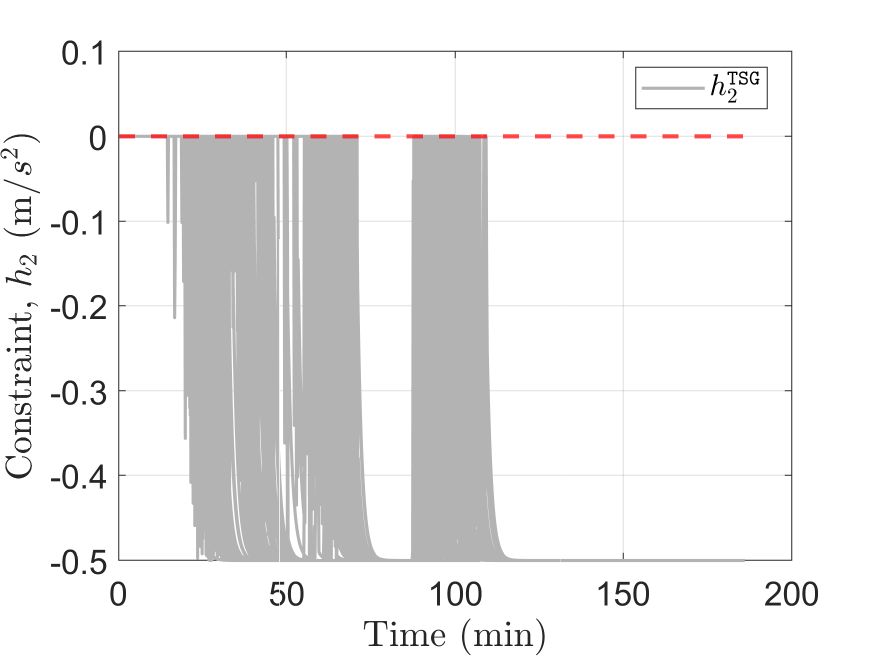}}\;
  \sidesubfloat[]{\label{fig:constraint_h3}\includegraphics[width = 1.85 in]{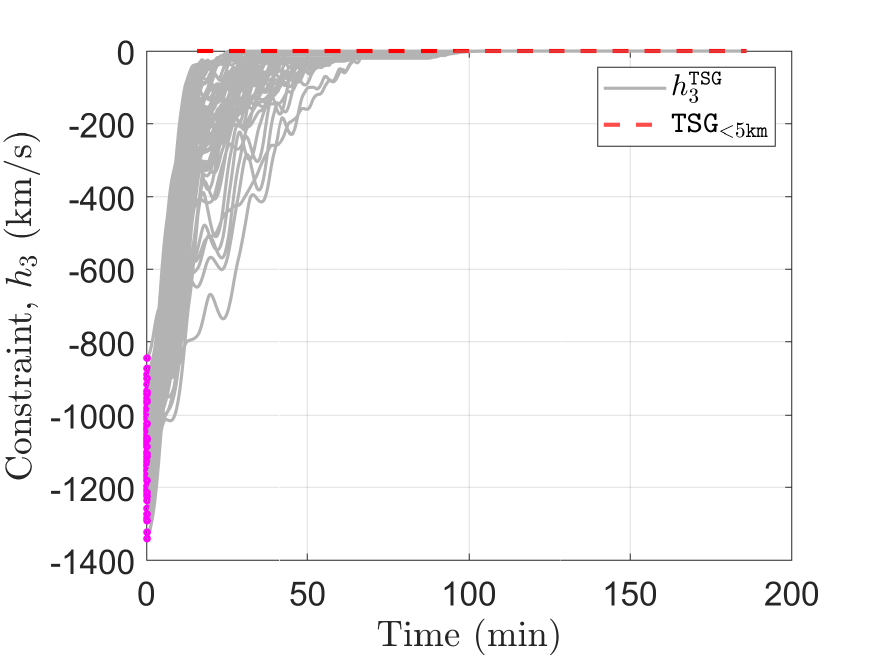}}\;
  \caption{Time histories of a) LoS cone angle constraint $h_1$; b) thrust constraint $h_2$; c) approach velocity constraint $h_3$.}\label{fig:constraints}
\end{figure}

\floatsetup[figure]{style=plain}
\captionsetup[subfigure]{labelfont=bf,textfont=normalfont}
\begin{figure}
  \sidesubfloat[]{\label{fig:control_input_history}\includegraphics[width=0.41\linewidth]{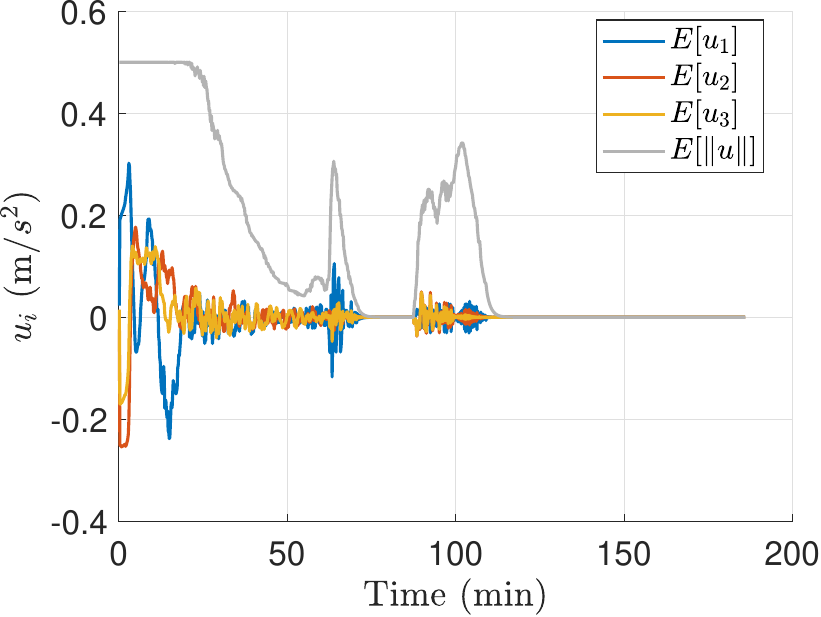}}\;
  \sidesubfloat[]{\label{fig:timeshift_history}\includegraphics[width=0.43\linewidth]{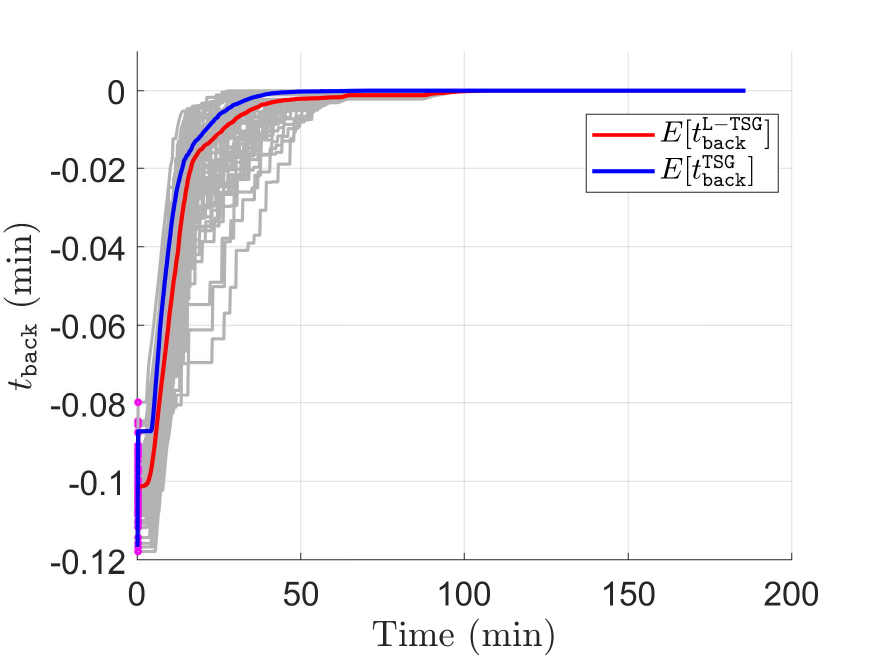}}\;
  \caption{a) Control input history and b) time shift history throughout the mission.}\label{fig:control_timeshift}
\end{figure}

\subsection{Simulation in Low Earth Orbit}\label{subsec3}
% Figure~\ref{fig:trajectory_3d_MC} shows the Deputy spacecraft's progressive approach and successful RD with the Chief spacecraft. Particularly, Fig. \ref{fig:trajectory_3d_MC} demonstrates the DL-TSG in guiding the RD mission by adjusting the Deputy's virtual target within the VNB frame. This control strategy, leveraging the DL-TSG, ensures mission completion without violating operational constraints.
% Particularly, Fig. \ref{fig:trajectory_3d_MC} demonstrates the DL-TSG in guiding the RD mission by adjusting the Deputy's virtual target within the VNB frame. This control strategy, leveraging the DL-TSG, ensures mission completion without violating operational constraints.

In this scenario, the Chief spacecraft orbits along the LEO of the International Space Station (ISS) and the Deputy spacecraft nominally starts about 50 km behind the Chief spacecraft in the same orbital track.

Table~\ref{tab:initial_condition} summarizes the estimated initial conditions of the Deputy spacecraft with respect to the Chief spacecraft in the ECI frame. This simulation employs Monte Carlo methods to introduce variations, where 100 different initial states of the Deputy spacecraft are tested.

Figure~\ref{fig:trajectory_3d_MC} illustrates the Deputy spacecraft trajectories in the VNB frame for 100 different initial conditions, marked as purple circles, using the L-TSG during RD missions. The L-TSG provides time-shifted Chief spacecraft trajectories, marked as magenta asterisks, to enforce various constraints for the Deputy spacecraft. For all simulations, the Deputy spacecraft with L-TSG is capable of completing the rendezvous with the Chief spacecraft, marked as a black circle, within the Line of Sight (LoS) cone, as shown in Fig.~\ref{fig:trajectory_3d_MC}.

%Figure~\ref{fig:rel_state_MC} shows the Deputy spacecraft position and velocity relative to the Chief spacecraft and to the virtual target in the ECI frame. The Deputy spacecraft becomes gradually closer to the Chief spacecraft and successfully achieves the RVD mission.
Figure~\ref{fig:rel_state_MC} shows the mean of the relative position and velocity of the Deputy spacecraft with respect to the Chief spacecraft and the virtual target, respectively, for two orbit periods in the ECI frame. In Fig.~\ref{fig:relposStateChief}, the Deputy spacecraft starts 50 km away from the Chief spacecraft, behind the Chief spacecraft along the orbital track, and approaches the Chief spacecraft. During the one orbit period, the mean of the Deputy spacecraft trajectories reaches close to the trajectory of the Chief spacecraft in Figs.~\ref{fig:relposStateChief},~\ref{fig:relvelStateChief}. Then, the mean of the relative position and velocity of the Deputy spacecraft with respect to the Chief spacecraft maintains near zero after one orbit period, which means the Deputy spacecraft stays in the vicinity of the Chief spacecraft. Figures~\ref{fig:relposStateTarget} and \ref{fig:relvelStateTarget} provide the mean of the virtual target trajectories that are the closest target reference, along the reference orbital track, to the nominal closed-loop system to satisfy the constraints. Thus, the relative distance and velocity of the Deputy spacecraft with respect to the virtual target is smaller than that with respect to the Chief. 
% The effective management of these constraints can be attributed to the enhanced prediction capabilities of the DL-TSG, which leverages the LSTM network's ability to process and remember complex patterns over time, providing accurate time shift prediction that enforces mission constraints.

% Performance of the constraint enforcement under the DL-TSG during the RD mission is outlined in Fig.~\ref{fig:constraint_h1},~\ref{fig:constraint_h2}, and \ref{fig:constraint_h3}. Specifically, Fig.~\ref{fig:constraint_h1} confirms the effective management of the Line of Sight (LoS) cone angle constraint $h_1$. Thrust constraint $h_2$ and the relative velocity constraint $h_3$ are presented in Fig.~\ref{fig:constraint_h2} and \ref{fig:constraint_h3}, respectively. Figure \ref{fig:constraint_h3} shows that our proposed control approach enforces the constraint during the docking phase when the Deputy spacecraft approaches within a 5km range of the Chief.

In Fig.~\ref{fig:constraints}, the time histories of three inequality constraints are presented from Monte Carlo simulations with perturbed initial conditions. Figure~\ref{fig:constraint_h1} provides the LoS constraint evolution during the RD mission, and the L-TSG is capable of enforcing the LoS cone angle constraint \(h_{1}\) for the perturbed initial conditions. Figure~\ref{fig:constraint_h3} shows the time histories of the relative velocity constraint \(h_{3}\) given varying initial relative velocities, and L-TSG is able to handle this constraint during the RD missions. Note that the thrust limit is applied by the saturation function in Eq.~\eqref{eq:saturated_control_input}, instead of L-TSG, but it narrows the feasible set of the time shift by applying the limited thrust in the validation process. Monte Carlo simulation results show that the L-TSG effectively enforces these constraints, as the LSTM cell in the L-TSG captures complex patterns to estimate the optimal time shift parameter.

%Figure~\ref{fig:timeshift_history} illustrates the time shift parameter $t_{\tt{back}}$'s adaptation over the mission. Starting from the initial permissible value ${t_{{\tt{back}}}(0)}$, the parameter increases using the prediction model's time shift prediction. This increment continues until the Deputy is positioned optimally to rendezvous with the Chief, at which point $t_{\tt{back}}$ is reduced to zero, allowing the Deputy to complete the approach under the nominal control scheme without violating any constraints.

Figure~\ref{fig:timeshift_history} illustrates the trajectories of the time shift parameter $t_{\tt{back}}$ as a function of time in a Monte Carlo campaign. Starting with a valid initial time shift ${t_{{\tt{back}}}(0)}$, the parameter gradually increases to zero using L-TSG. This indicates that the Deputy spacecraft follows the virtual target's trajectory, without constraint violations, and eventually stabilizes near the Chief spacecraft. The deep learning model's training process uses the MSE term to drive the time shift estimate toward the optimal solution, while the MSReLU term prevents estimates that would lead to constraint violations. As a result, L-TSG produces maximum admissible time shifts that guarantee safety, given the standard LQ controller as a nominal closed-loop system.

% Initial condition
\begin{table}[b]
\caption{Expected initial state of the Deputy relative to the Chief, e.g., \( \mathbb{E}\big[X_d(t_{0}) - X_c(t_{0})\big] \).}
\label{tab:initial_condition_molniya}
\centering
\begin{tabular}{cccccc}
\hline
\( \delta x_0 \) [km] & \( \delta y_0 \) [km] & \( \delta z_0 \) [km] & \( \delta{\dot x_0} \) [km/s] & \( \delta{\dot y_0} \) [km/s] & \( \delta{\dot z_0} \) [km/s] \\
\hline
-9.7168 & -0.3110 & 0.5869 & 0.0014 & -0.0035 & -0.0068 \\
\hline
\end{tabular}
\end{table}

\begin{figure}[t]
\centering
\includegraphics[width=0.8\linewidth]{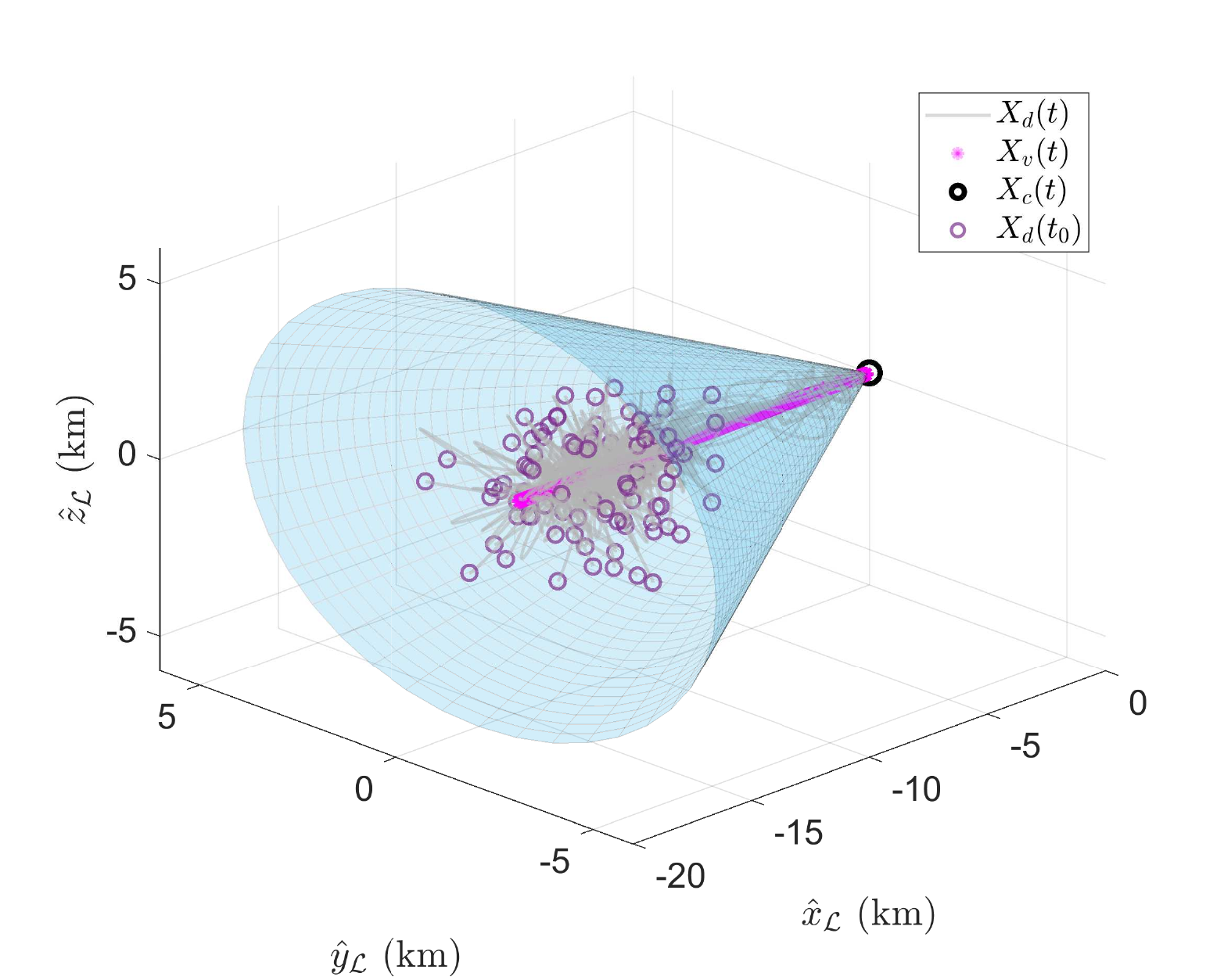}
\caption{Trajectories displayed in the local VNB frame: the deputy spacecraft’s path (blue line) closely follows the virtual target’s trajectory (magenta asterisk) in a highly eccentric orbit.}
\label{fig:trajectory_3d_MC_molniya}
\end{figure}

\floatsetup[figure]{style=plain}
\captionsetup[subfigure]{labelfont=bf,textfont=normalfont}
\begin{figure}
  \sidesubfloat[]{\label{fig:relposStateChief_molniya}\includegraphics[width = 2.5 in]{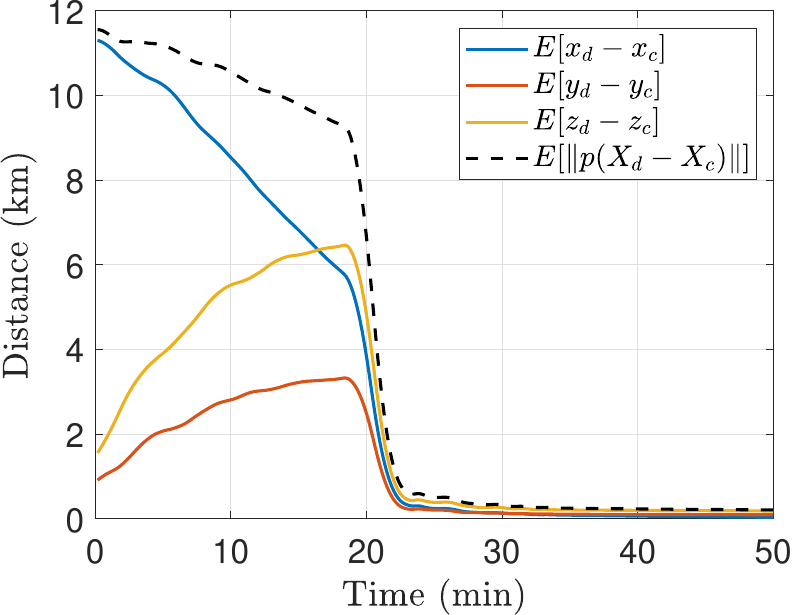}}\;%
  \sidesubfloat[]{\label{fig:relvelStateChief_molniya}\includegraphics[width = 2.5 in]{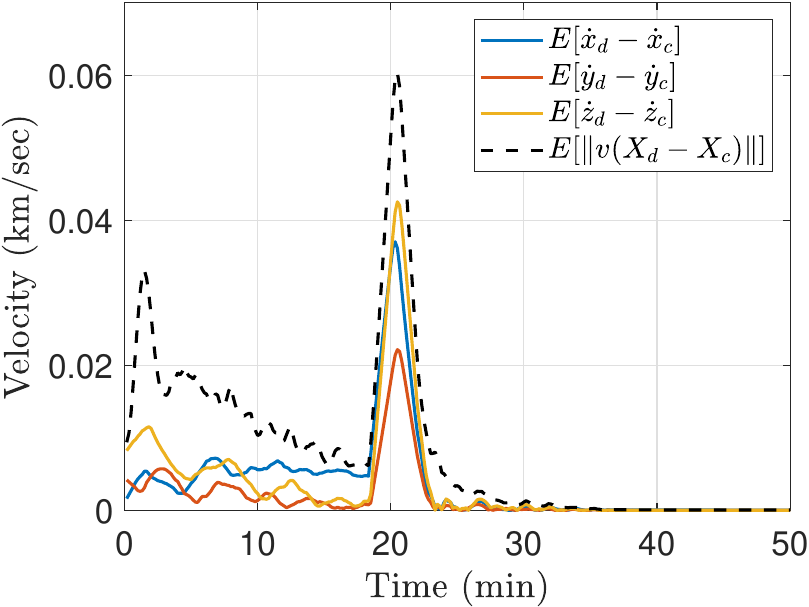}}\\
  \sidesubfloat[]{\label{fig:relposStateTarget_molniya}\includegraphics[width = 2.5 in]{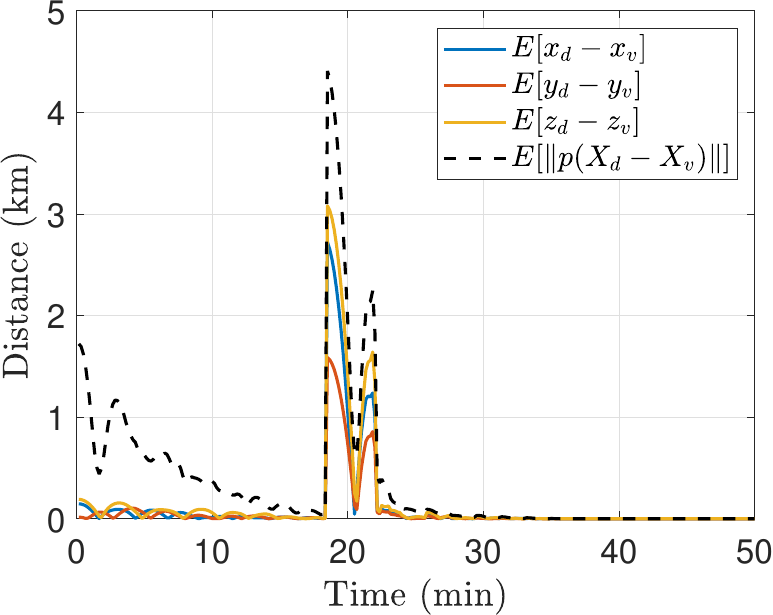}}\;
  \sidesubfloat[]{\label{fig:relvelStateTarget_molniya}\includegraphics[width = 2.5 in]{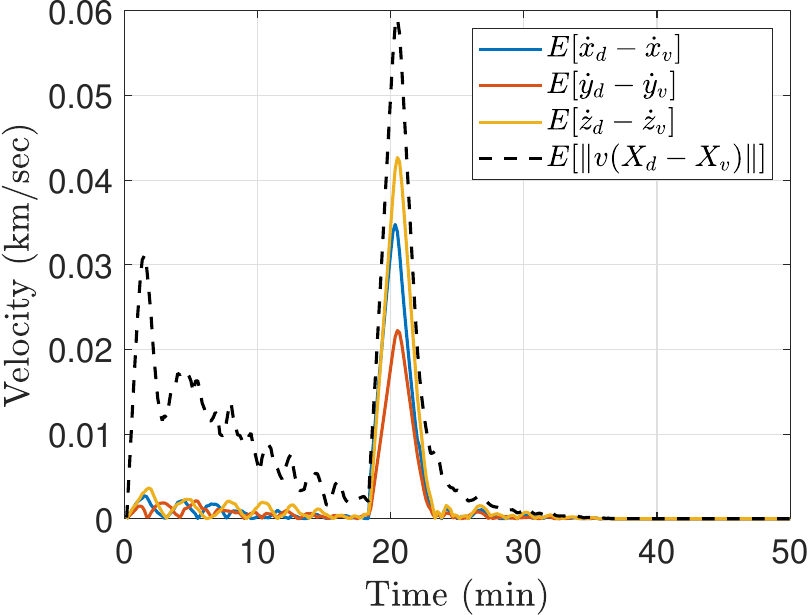}}\;
  \caption{Time histories of relative a) position; b) velocity to the Chief spacecraft; c) position; d) velocity to the virtual target, in an elliptical orbit.}\label{fig:rel_state_MC_molniya}
\end{figure}

\floatsetup[figure]{style=plain}
\captionsetup[subfigure]{labelfont=bf,textfont=normalfont}
\begin{figure}
  \sidesubfloat[]{\label{fig:constraint_h1_molniya}\includegraphics[width = 1.85 in]{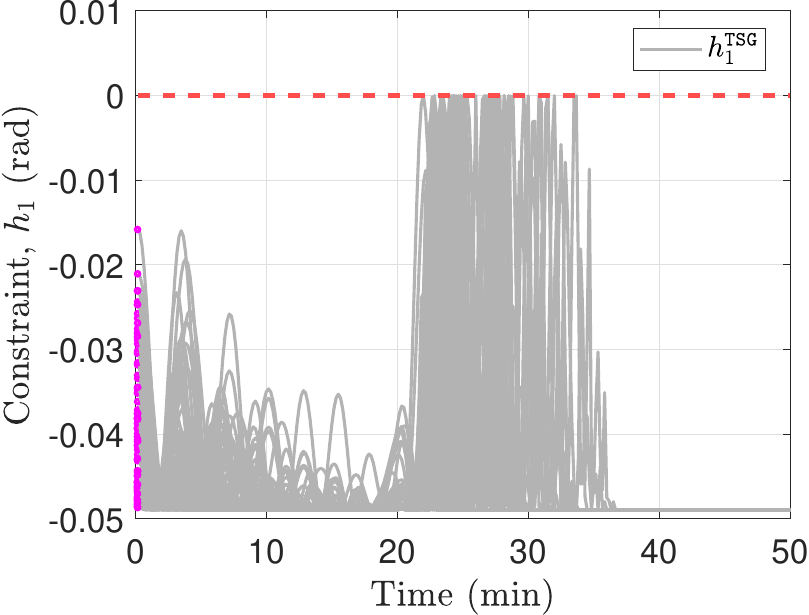}}\;%
  \sidesubfloat[]{\label{fig:constraint_h2_molniya}\includegraphics[width = 1.85 in]{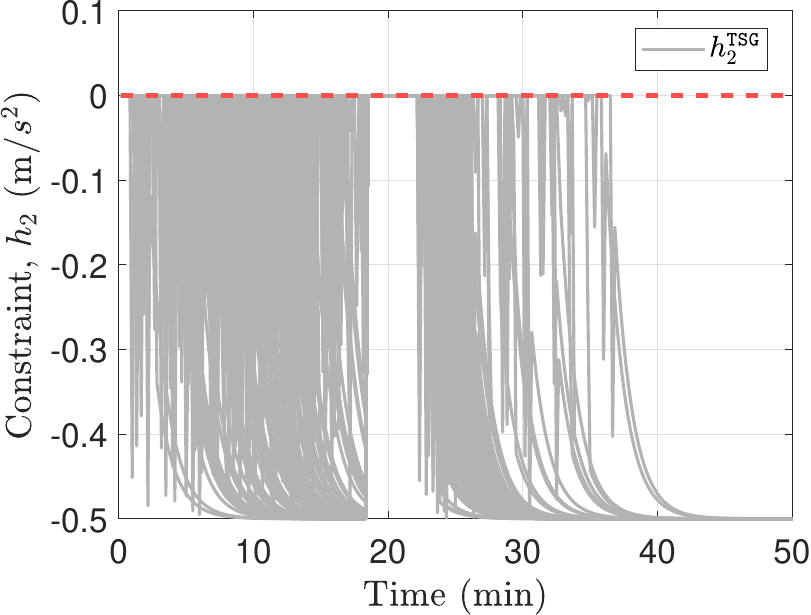}}\;
  \sidesubfloat[]{\label{fig:constraint_h3_molniya}\includegraphics[width = 1.85 in]{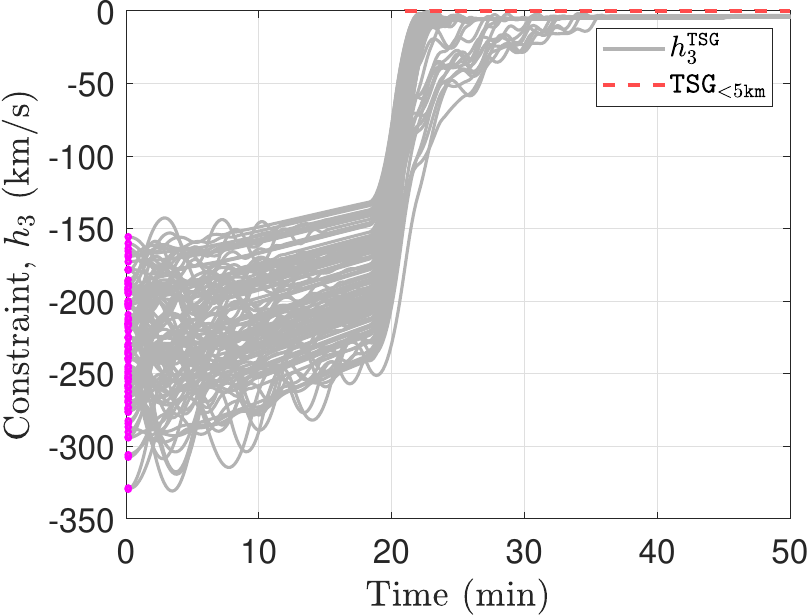}}\;
  \caption{Time histories of a) LoS cone angle constraint $h_1$; b) thrust constraint $h_2$; c) approach velocity constraint $h_3$ for Monte Carlo simulations in a high-eccentricity orbit.}\label{fig:constraints_molniya}
\end{figure}

\floatsetup[figure]{style=plain}
\captionsetup[subfigure]{labelfont=bf,textfont=normalfont}
\begin{figure}
  \sidesubfloat[]{\label{fig:control_input_history_molniya}\includegraphics[width=0.41\linewidth]{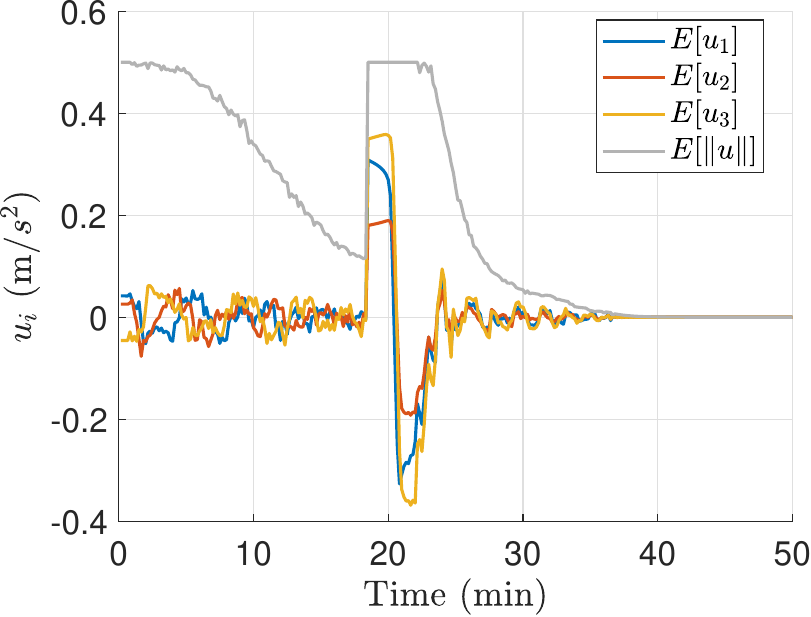}}\;
  \sidesubfloat[]{\label{fig:timeshift_history_molniya}\includegraphics[width=0.40\linewidth]{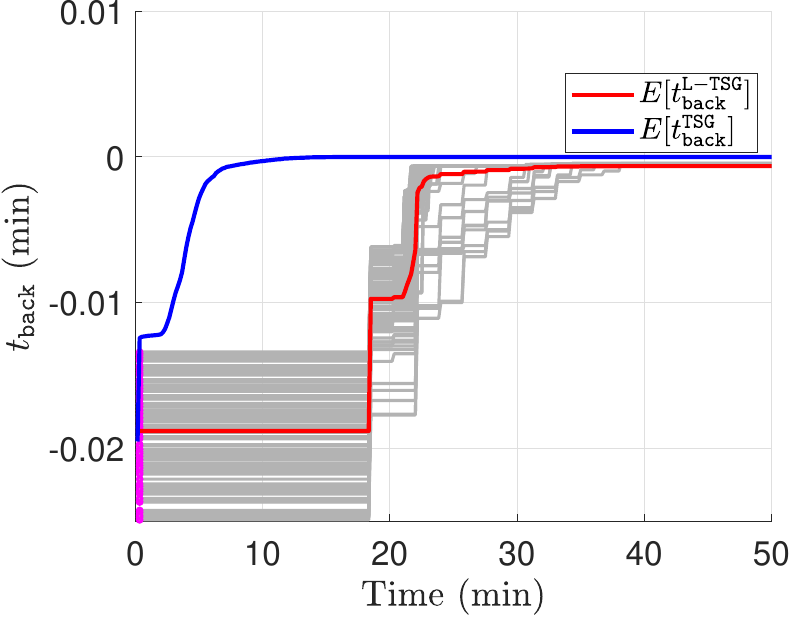}}\;
  \caption{a) Control input and b) time shift history throughout the mission in a Molniya orbit.}\label{fig:control_timeshift_molniya}
\end{figure}

\begin{table}[t]
\caption{Performance comparisons for spacecraft rendezvous missions}
\label{tab:comp_time_delta_v}
\centering
\begin{tabular}{c|ccc|ccc}
\hline \hline
Method  & \multicolumn{3}{c|}{LEO} & \multicolumn{3}{c}{Molniya Orbit} \\
        & Avg. Time (s) & Worst-case (s) & $\Delta V$ (km/s) & Avg. Time (s) & Worst-case (s) & $\Delta V$ (km/s) \\
\hline
TSG & 0.1476 & 0.2057 & 1.2939 & 0.0852 & 0.1665 & 0.3558 \\
L-TSG & 0.0233 & 0.0328 & 2.2477 & 0.0256 & 0.0323 & 0.9889 \\

\hline \hline
\end{tabular}
\end{table}

\subsection{Simulation in Molniya Orbit}\label{subsec4}
We next apply L-TSG to a more challenging scenario: the Molniya orbit. The Molniya orbit is characterized by its high eccentricity, providing a different set of challenges for the control system compared to the nearly circular orbit of the Crew-3 mission.

In this scenario, the Chief spacecraft orbits along the highly elliptical Molniya orbit, while the Deputy spacecraft starts approximately 10 km behind the Chief. The initial conditions of the Deputy spacecraft relative to the Chief are summarized in Table~\ref{tab:initial_condition_molniya}. As in the Crew-3 mission, we conducted Monte Carlo simulations with 100 different initial states of the Deputy spacecraft to introduce variations and assess the robustness of the control system.

Figure~\ref{fig:trajectory_3d_MC_molniya} illustrates the Deputy spacecraft's trajectories in the VNB frame for these 100 initial conditions. The L-TSG provides time-shifted Chief spacecraft trajectories (marked by magenta asterisks) to enforce various constraints on the Deputy spacecraft. In all cases, the Deputy spacecraft successfully completed the rendezvous with the Chief spacecraft (marked by a black circle) within the Line of Sight (LoS) cone, demonstrating L-TSG's capability in handling the high-eccentricity orbit.

Figure~\ref{fig:rel_state_MC_molniya} shows the mean relative position and velocity between the Deputy spacecraft and both the Chief spacecraft and the virtual target over two orbital periods in the ECI frame. In Fig.~\ref{fig:relposStateChief_molniya}, the Deputy spacecraft starts 10 km behind the Chief and gradually approaches it. By the end of the orbital period, the Deputy spacecraft reaches and remains close to the Chief spacecraft’s trajectory (as seen in Figs.~\ref{fig:relposStateChief_molniya} and \ref{fig:relvelStateChief_molniya}). Similarly, the relative position and velocity with respect to the virtual target (Figs.~\ref{fig:relposStateTarget_molniya} and \ref{fig:relvelStateTarget_molniya}) show that the virtual target provides an effective reference for ensuring constraint satisfaction.

In Fig.~\ref{fig:constraints_molniya}, the time histories of the three main inequality constraints—LoS cone angle (\(h_1\)), thrust (\(h_2\)), and approach velocity (\(h_3\))—are shown for the Monte Carlo simulations with perturbed initial conditions. Fig.~\ref{fig:constraint_h1_molniya} illustrates that L-TSG successfully enforced the LoS constraint throughout the simulation. Fig.~\ref{fig:constraint_h3_molniya} demonstrates that the relative velocity constraint was also effectively managed. Although the thrust constraint (\(h_2\)) is enforced through saturation, the L-TSG manages the time shift to stay within the feasible set of constraints, as shown in Fig.~\ref{fig:constraint_h2_molniya}. These results indicate that the L-TSG handled this high-eccentricity, challenging orbit well. It dynamically adjusted the time shift in response to rapidly varying orbital conditions, ensuring that the Deputy spacecraft remained within safe operational limits.

The success of the L-TSG in this highly elliptic orbit can be attributed to several key factors. First, the high eccentricity of the Molniya orbit introduces significant variations in both velocity and distance during each orbital pass, which makes trajectory prediction and constraint enforcement more challenging. The LSTM-based deep learning model within L-TSG was able to capture the temporal dependencies and nonlinearities in these rapidly changing conditions, allowing for accurate time shift predictions. Additionally, the virtual target, calculated using the time shift, acted as a dynamic reference point that effectively guided the Deputy spacecraft to adjust its trajectory, even under these difficult conditions. This ensured that the spacecraft could maintain safe relative distances and velocities despite the large variations introduced by the orbit’s eccentricity.

Finally, Fig.~\ref{fig:timeshift_history_molniya} displays the evolution of the time shift parameter \(t_{\tt{back}}\) during the mission. The parameter starts at an initial value \(t_{\tt{back}}(0)\) until the window size of 100 for the LSTM-based model is filled. Once the window is filled, the time shift gradually approaches zero as the Deputy spacecraft stabilizes near the Chief. This gradual adjustment is particularly significant in a highly elliptical orbit like Molniya’s, where abrupt changes in velocity and distance are expected at different points along the orbit. The L-TSG efficiently handles these changes by incrementally updating the time shift, allowing the Deputy spacecraft to track the virtual target without constraint violations. The learning-based model, guided by the MSE and MSReLU terms, was able to predict time shifts that accounted for the eccentric orbit’s challenging dynamics, avoiding solutions that would have led to constraint violations. This capability allowed the L-TSG to achieve a successful rendezvous with the Chief spacecraft despite the complexities of the orbit.

Compared to the conventional TSG, Table~\ref{tab:comp_time_delta_v} shows that the L-TSG significantly reduces the onboard computation time for time shift parameter prediction. By using a prediction of the LSTM network, the L-TSG can generate the corresponding virtual target by checking the validity of the time shift, reducing computation times from 0.1476 s to 0.0233 s in LEO and from 0.0852 s to 0.0256 s in the Molniya orbit. When the Deputy spacecraft is far from the Chief spacecraft, the LSTM model updates a more accurate time shift than the prediction when it is near the Chief spacecraft. Depending on the relative distance, we replace an LSTM model, according to the PA-SW approach, to address this issue. Before the model transitions or mission completion, L-TSG potentially uses the backup plan of using the conventional TSG, but it can still contribute to computation time reduction for average and worst-case because of the narrowed bounds of the time shift set.

\section{Conclusion} \label{sec:conclusion}
This study provides and demonstrates a novel framework that integrates a Time Shift Governor (TSG) with a learning-based model, called L-TSG, to accelerate the performance of TSG during rendezvous and docking (RD) missions in the Two-Body problem setting. By incorporating the learning-based model in our Monte Carlo campaign, L-TSG successfully avoided 95.4\% of iterative processes in the bisection-based TSG during the Crew-3 mission and 95.1\% during the Molniya orbit scenario. 
It showcases its capability to enforce state and control constraints during close rendezvous, including approach direction, relative velocity, and thrust limits. A target reference computed from L-TSG guides the closed-loop system to avoid constraint violations via a time-shifted state trajectory of the Chief spacecraft. The Deputy spacecraft conducts the RD mission until achieving the target reference, which aligns with the Chief spacecraft. 

\bibliography{reference}

\end{document}